\documentclass{aa}

\usepackage{graphicx,natbib}
\usepackage{amsmath} 
\usepackage{ulem} 
\bibpunct{(}{)}{;}{a}{}{,}

\def\ergscm{erg~s$^{-1}$~cm$^{-2}$}

\def\arcmin{\hbox{$^\prime$}}

\def\aj{AJ}
\def\apj{ApJ}
\def\apss{Astroph.Sp.Sci.}
\def\aap{A\&A}

\begin{document}

\title{INTEGRAL/IBIS 7-year All-Sky Hard X-Ray Survey\thanks{Based on observations with INTEGRAL, an ESA project with the 
instruments and science data centre funded by ESA member states
(especially the PI countries: Denmark, France, Germany, Italy,
Switzerland, Spain), Czech Republic and Poland, and with the
participation of Russia and the USA.}\\ Part I: Image Reconstruction.}

\author{R.~Krivonos\inst{1,2}, M.~Revnivtsev\inst{2,3}, S.~Tsygankov\inst{1,2}, S.~Sazonov\inst{2}, A.~Vikhlinin\inst{2,4},
M.~Pavlinsky\inst{2}, E.~Churazov\inst{1,2} \and R.~Sunyaev\inst{1,2}}

\institute{
              Max-Planck-Institute f\"ur Astrophysik,
              Karl-Schwarzschild-Str. 1, D-85740 Garching bei M\"unchen,
              Germany
\and
              Space Research Institute, Russian Academy of Sciences,
              Profsoyuznaya 84/32, 117997 Moscow, Russia
\and 
              Excellence Universe Cluster, Munich Technical University, Boltzman-Str. 2, D-85748 Garching bei M\"unchen, Germany
\and
              Harvard-Smithsonian Center for Astrophysics, 60 Garden Street, Cambridge, MA 02138, USA
            }
\authorrunning{Krivonos et al.}

\abstract{This paper is the first in a series devoted to the hard
  X-ray whole sky survey performed by the INTEGRAL observatory over
  seven years. Here we present an improved method for image
  reconstruction with the IBIS coded mask telescope. The main
  improvements are related to the suppression of systematic effects
  which strongly limit sensitivity in the region of the Galactic Plane
  (GP), especially in the crowded field of the Galactic Center
  (GC). We extended the IBIS/ISGRI background model to take into
  account the Galactic Ridge X-ray Emission (GRXE). To suppress
  residual systematic artifacts on a reconstructed sky image we
  applied nonparametric sky image filtering based on wavelet
  decomposition. The implemented modifications of the sky
  reconstruction method decrease the systematic noise in the
  $\sim20$~Ms deep field of GC by $\sim44\%$, and practically remove
  it from the high-latitude sky images.  New observational data sets,
  along with an improved reconstruction algorithm, allow us to conduct
  the hard X-ray survey with the best currently available minimal
  sensitivity $3.7\times10^{-12}$~\ergscm ~$\sim0.26$~mCrab in the
  17-60~keV band at a $5\sigma$ detection level. The survey covers
  $90\%$ of the sky down to the flux limit of
  $6.2\times10^{-11}$~\ergscm ~($\sim4.32$~mCrab) and $10\%$ of the
  sky area down to the flux limit of $8.6\times10^{-12}$~\ergscm
  ~($\sim0.60$~mCrab).

} \maketitle

\section{Introduction}

Since its launch in October 2002, the INTEGRAL observatory
\citep{integral} has gathered a huge observational data set allowing
us to perform the most sensitive hard X-ray survey to date.  The main
scientific results and source catalogues have been reported in many
relevant papers concerning partial sky coverage
\citep[e.g.][]{revetal03a,molkov2004,krietal05,revetal06a,birdI,birdII,birdIII,bassani06,bazzano06}
and full sky studies \citep{krietal07b,sazonov07,beckmann09,birdIV}.

Recently, great progress in surveying the hard X-ray sky was achieved
with the Burst Alert Telescope \citep[BAT;][]{bat} at the \textit{Swift}
observatory \citep{swift}. The \textit{Swift}/BAT survey provides very
homogeneous sky coverage in the 15-195~keV energy band with a current
maximum sensitivity of $2.2\times10^{-11}$\ergscm. The distribution of
survey sensitivity peaks in the extragalactic sky. The survey results
and source catalogues have been reported in papers by \cite{bat22} and
\cite{palermo36}. As seen from the large sample of detected Active
Galactic Nuclei (AGNs), the results of the \textit{Swift}/BAT survey are
very valuable for extragalactic studies. However, due to the
relatively poor angular resolution of the instrument, its capabilities
in the Galactic plane and especially in the Galactic Center regions
are limited. On the other hand, the sky coverage by the
\textit{Swift}/BAT survey is nearly uniform, therefore only a small
fraction of its total operational time was devoted to observations of
the Galaxy.

Contrary to \textit{Swift}, the INTEGRAL observatory provides an all-sky survey
with exposure more concentrated in the Galactic Plane, having a typical
limiting flux of less than $1.43\times10^{-11}$\ergscm\ (1 mCrab) in
the working energy range 17--60~keV. With an angular resolution almost
twice as good as \textit{Swift}/BAT, one can effectively disentangle
sources in such crowded regions as the Galactic Center. This makes
the \textit{Swift}/BAT and INTEGRAL surveys complementary to each other.

INTEGRAL has already accumulated a lot of exposure time in the
direction of the Galactic Plane with a maximum of $\sim$20~Ms of
nominal time in the direction of the Galactic Center.  However, the
growing exposure time devoted to the Plane and the Center of the
Galaxy is not reflected by a corresponding increase in survey
sensitivity. Observations in these regions are strongly affected by
the systematics related to the crowded field of the GC and strong
Galactic X-ray background radiation.

In this work we address the question of improving the sensitivity of
the ongoing INTEGRAL hard X-ray survey. In section
(\ref{section:method}) we discuss several aspects of the sky
reconstruction method of the IBIS coded-mask telescope
\citep{ibis}. In section (\ref{section:ridge}) we implement Galactic
background corrections to the sky reconstruction method. Section
(\ref{section:method2}) introduces a modified sky reconstruction
method with sky image filtering procedure based on
\textit{\`a~trous}\/ wavelet decomposition. The properties of the
resulting all-sky survey are presented in section
(\ref{section:survey}).

\begin{figure}\centerline{
 \includegraphics[width=0.5\textwidth]{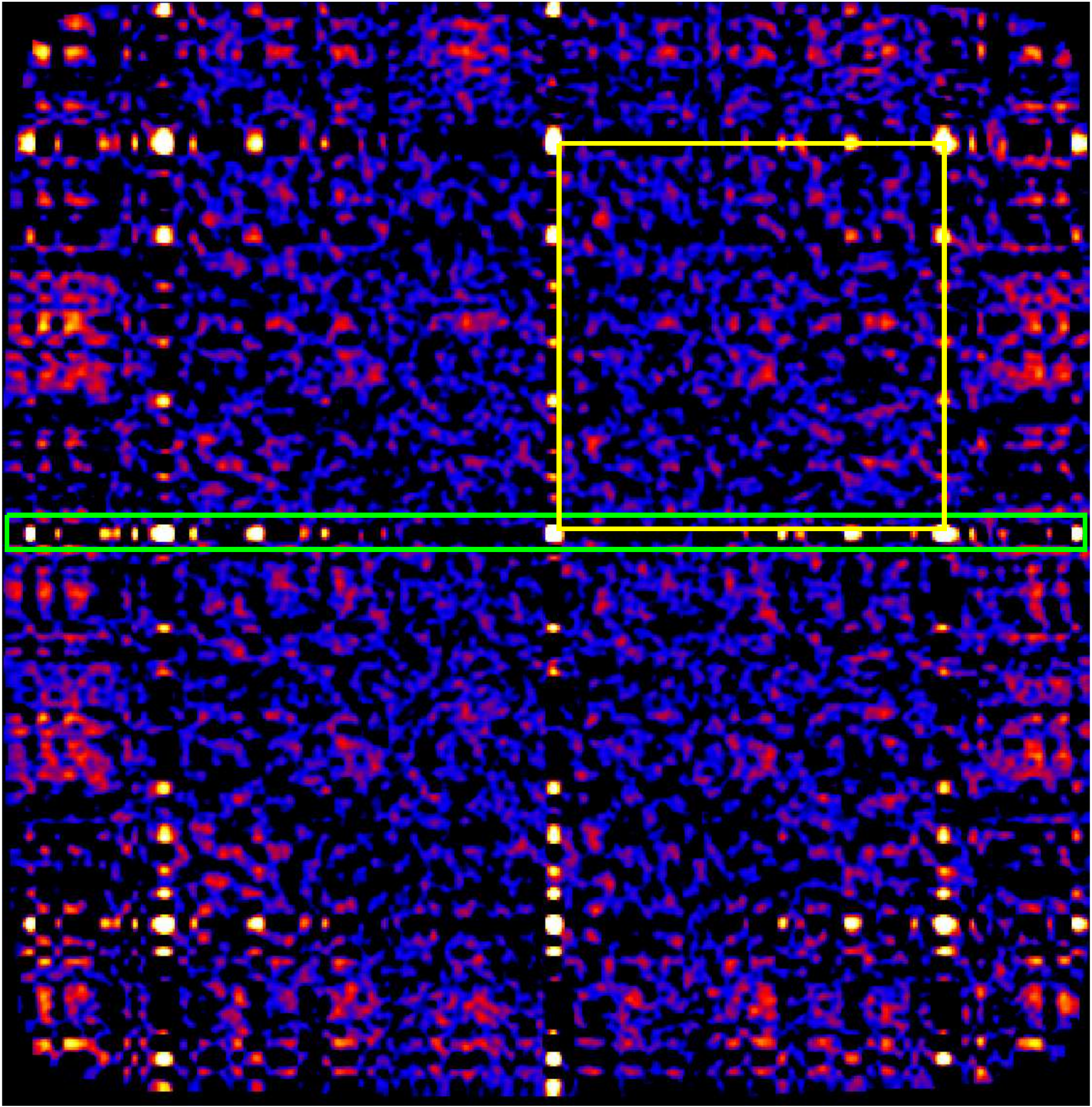}}
\caption{Sky image of a $2.7$~ks observation of the bright source
  Crab Nebula. The source is in the center of the field of view. The
  periodic mask pattern leads to many false sources on the
  reconstructed sky. The most significant false peaks are located in
  vertexes of a $\sim10.5^{\circ}$ square around the source, oriented
  in detector coordinates as illustrated by the yellow square.  The
  image is shown in significance with a squared root color map ranging
  from 0 to 25. Such color schemes are used for all the sky images in
  this paper in order to emphasize sky background variations. The
  black and blue colors correspond to pixel values from 0 to 2. The
  red pixels have values of around $5$. The yellow to white color
  transition corresponds to $15$ and more.  Fig.~\ref{fig:crab:psf}
  demonstrates an image profile extracted from the $1^{\circ}\times
  29^{\circ}$ green region.}\label{fig:crab:sky}
\end{figure}

Throughout the article the exposure will be expressed taking into
account instrumental dead time, i.e. showing the effective exposure
time, rather than the total exposure.

\section{General sky reconstruction method}
\label{section:method}

IBIS is a coded aperture imaging telescope. The sky is projected on to
the detector plane through the transparent and opaque elements of the
mask mounted above the detector plane. Generally, the sky
reconstruction is based on the deconvolution of the detector image
with a known mask pattern. We (EC) developed IBIS/ISGRI sky
reconstruction method and partially described it in our previous
publications \citep{revetal04,krietal05,krietal07a,krietal07b}. The
basic idea we used is presented in \cite{fenimore81} and
\cite{skinner87}. For the standard IBIS/ISGRI analysis we refer reader
to the paper by \cite{goldwurm03}. Here we outline only those steps
that are essential for the present study.

\begin{figure}\centerline{
 \includegraphics[width=0.5\textwidth]{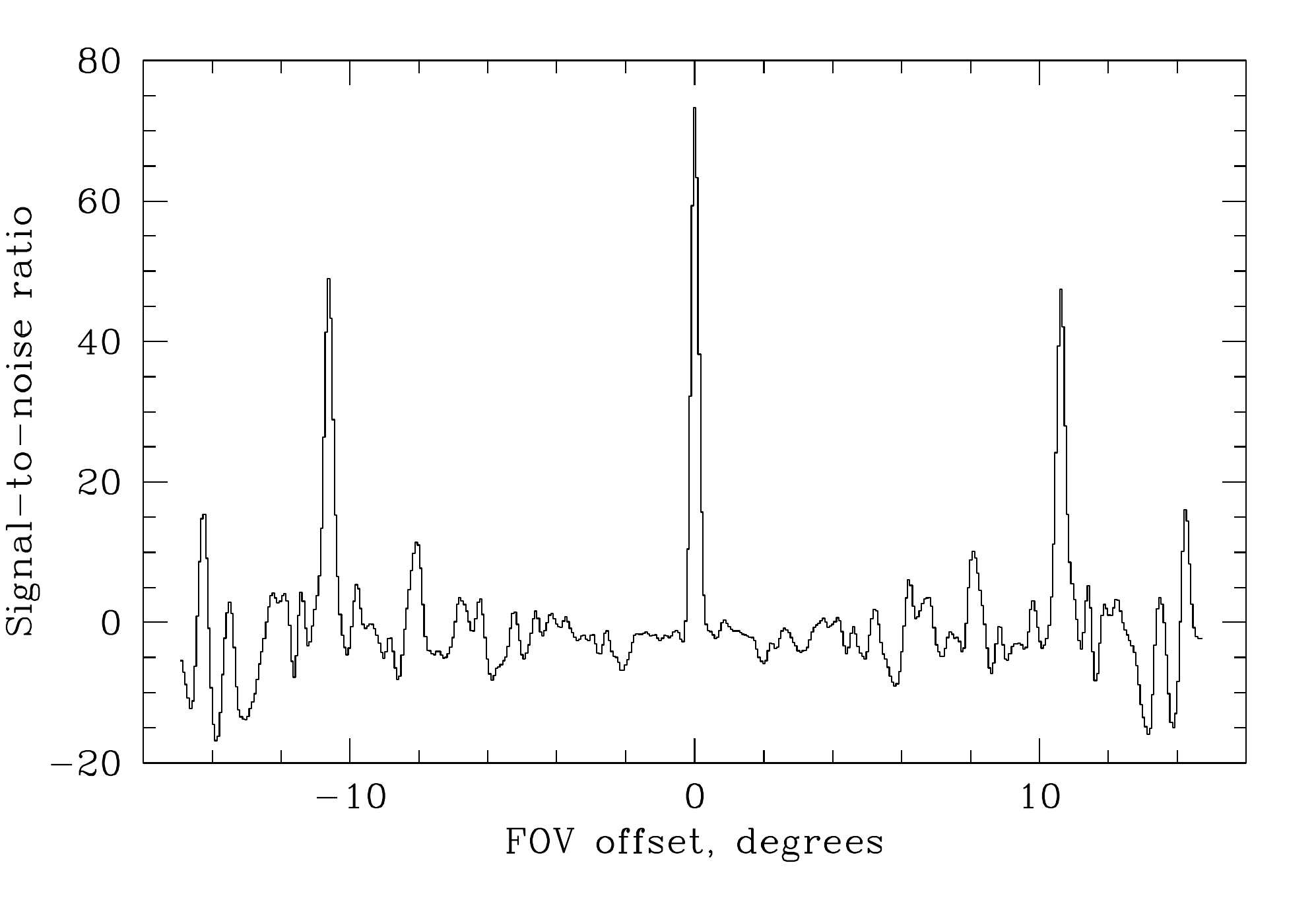}}
\caption{Image profile extracted from the green region in
  Fig.~\ref{fig:crab:sky}. The profile demonstrates the response of the
  IBIS/ISGRI imaging system to the point source.}\label{fig:crab:psf}
\end{figure}

\subsection{IBIS telescope coding aperture}
\label{section:mask}

The quality of the reconstructed sky image directly depends on our
understanding of the coding procedure. For example, the mask
supporting structure can significantly reduce the number of low energy
photons passed though the open mask elements. Other assembly elements
like screws, plates and glue strips attaching the IBIS mask to the
supporting structure also block the incoming photons modifying the
shadow cast by a point source on the detector.

In our package we have implemented the best known configuration of all
the known elements of the telescope, which make contributions to the
shadowgram. We could not find the exact size of the assembly
screws which attach the mask, plates and glue strips. The approximate
parameters of these elements were found by comparing real and model
detector shadowgrams illuminated by a strong source at different
azimuthal angles.

\subsection{Detector exposure}
\label{section:detector}

In general the detector image, produced when observing $M$ point
sources with an underlying flat sky background, can be represented by
the superposition of shadow patterns of sources (``pixel
illumination fraction'', $PIF$)  and the detector background map $B$:
\begin{equation}
D =  \sum_{i=0}^{M} f_iPIF_i + k_{B}B,
\label{eq:detcnts1}
\end{equation}     
where $f_{i}$ is the source flux, $B=B_{CXB}+B_{det}$ represents a
detector background map, containing photon counts from Cosmic X-ray
Background (CXB), and detector instrumental noise. We assume, that the
detector illumination by CXB and intrinsic ISGRI background have a
similar pattern, and can be merged into the background map $B$. The
last is estimated by an accumulation detector image over a large
number of observations without strong sources in the field of
view. Obviously, background map $B$ should contain the current
detector background pattern due to the long-term variation of
background environment related to the Sun and Cosmic rays
\citep{lebrun05}. For a given observation, we use a background map
accumulated during the nearest set of extragalactic
observations. Typically, we construct a new background map for every
$\sim50-70$ spacecraft orbits (150-200 days).

In the case when two sources located close to each other (at
separation comparable with IBIS/ISGRI angular resolution), the direct
solution of Eq.~\ref{eq:detcnts1} for flux $f_i$ can be unstable
giving results with infinite errors. In other words, the detector
count rate in a given sky direction can be explained by two (or more)
sources having any absolute flux. 

\subsection{Replicated mask pattern}
\label{section:mura}

IBIS mask has replicated patterns (see \citealt{reglero01},
\citealt{ibis}, \citealt{goldwurm03}). This pattern of the mask has an
advantage because, ideally, it has a much narrower point spread
function and flat side lobes in the central part of the reconstructed
image \citep{fenimore78}. But at the same time it causes serious
complications due to the presence of the very significant side peaks
of the point spread function. This means that a simple deconvolution
algorithm sees ``ghost'' sources at certain sky positions (see
Fig.~\ref{fig:crab:sky} and \ref{fig:crab:psf}), related to the
position of the real source and the size of the replicated mask
pattern.

\begin{figure*}[t]
\includegraphics[width=\textwidth]{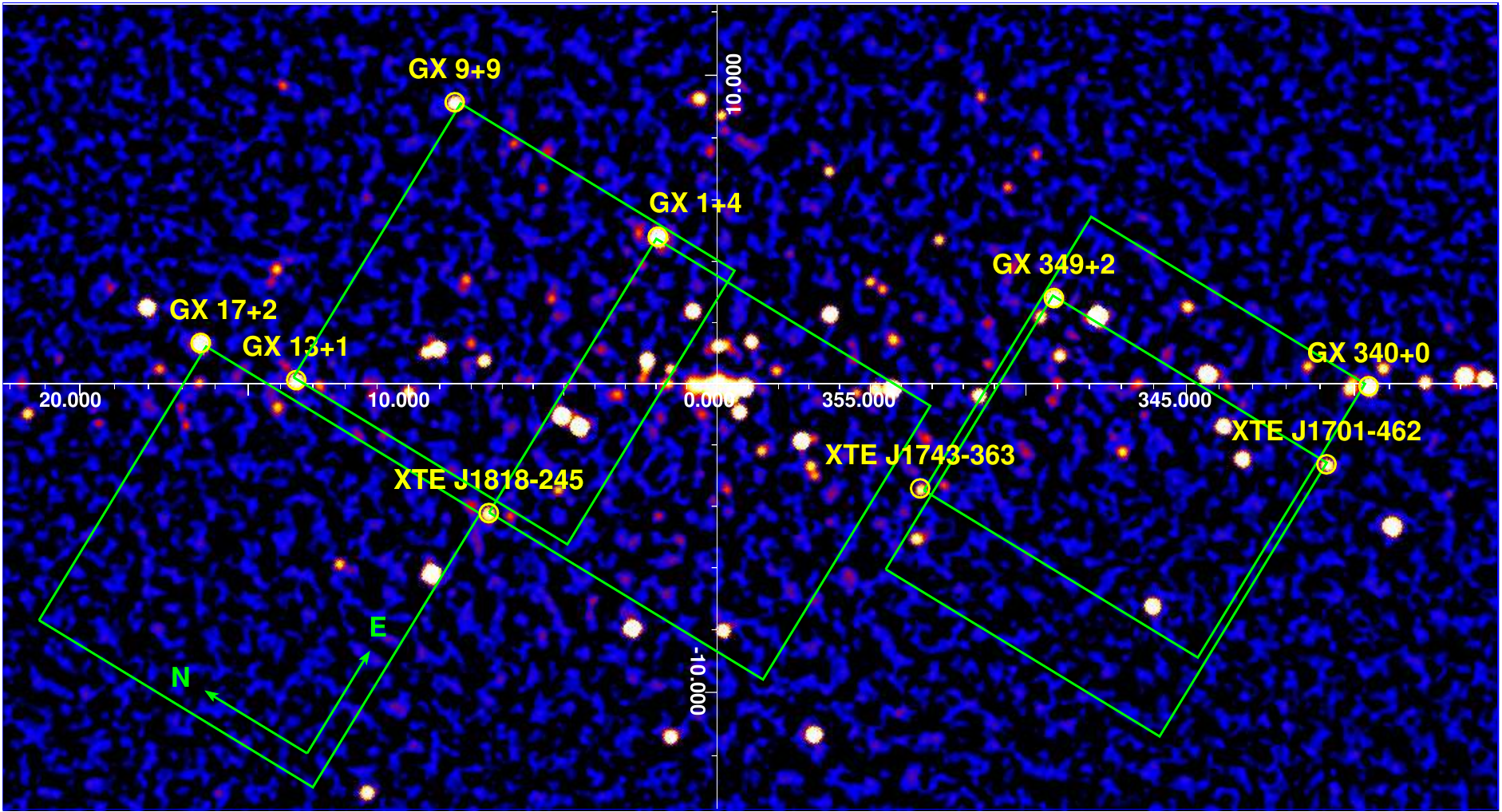}
\caption{INTEGRAL/IBIS hard X-ray ($17-60$~keV) map of the sky region
  around the Galactic Center. The green squares demonstrate the
  relative sky positions of false side peaks (``ghosts'', see
  Fig.\ref{fig:crab:sky}) of selected bright sources. Due to
  observational constrains IBIS FOV was mainly co-aligned with axes of
  the equatorial coordinate system (``N-E'' notation). As a result,
  there are a number of bright sources which permanently appeared in
  mutual ``ghost'' positions. This fact leads to an additional
  uncertainty of the source flux determination. The following source
  pairs can be affected by mutual flux interplay: GX~340+0 and
  XTE~J1743-363, XTE~J1701-462 and GX~349+2, GX~9+9 and GX~13+1, and
  the triplet GX~1+4, XTE~J1818-245 and GX~17+2.}\label{fig:ghosts}
\end{figure*}

The sources in a variety of sky positions within a field of view
create shadows with similar patterns, which causes uncertainty of
source flux determination (the direct solution of
Eq.~\ref{eq:detcnts1} is impossible). Unfortunately, this situation is
not rare in the crowded field of the Galactic Center, as shown in
Fig.~\ref{fig:ghosts}.

\subsection{Iterative removal of cataloged sources}
\label{section:iros}

In order to obtain a good quality all-sky map, suitable to searching
for new weak sources, the ghosts of known bright sources have to be
removed. This is done during the reconstruction of images of
individual observations (\textit{ScW}s) with an effective exposure of
$1-3$~ks. Instead of a blind search for bright sources in each
individual observation we use a catalogue of known sources to control
the removal of ghosts. Indeed, for many regions of sky the final map
is the result of stacking hundreds and thousands of individual
observations. Therefore, a relatively weak object, far too faint to be
detected in an individual observation, may appear as a very
significant source in the final map. Since the amplitude of the ghosts
scales with the intensity of the true source, it is clear that ghost
removal should be applied even to objects which are too faint to be
detected in individual science windows. For crowded fields (like the
Galactic Center region) this implies that ghosts of some 100 sources
should be removed in individual observations.

The whole procedure requires not only the list of sources to be
removed from the detector image, but also a sequence of
\textbf{I}terative \textbf{R}emoval \textbf{O}f \textbf{S}ources
(IROS, see \cite{goldwurm03,krietal05}). It is expected (and confirmed
by direct tests) that the brightest (most significant) objects have to
be removed first, since the source flux $f$ is evaluated assuming that
there is only one source in the field of view. The significance of the
source detection is evaluated by reconstructing an image prior to
iterative source removal and checking the fluxes at the positions of
cataloged sources. The list of objects ranked according to their
significance is then used as input for iterative source removal
procedure. This poses the problem of ranking weak sources, since their
flux (and ranking) is determined with a large uncertainty in the
individual science window (e.g. the flux from the source can be
negative). We made several tests with various ranking schemes for weak
(less than $3\sigma$ detection) sources, checking the RMS of the final
maps and fluxes of cataloged sources. The final scheme implemented in
our analysis uses the \textit{absolute} value of the source detection
significance to rank the order of source removal.

\subsection{ISGRI pixel filtering}
\label{section:pixels}
 
As described in \cite{krietal07b} the hot and dead ISGRI detector
pixels were screened from the analysis. This was done using quite
crude filtering criteria and some noisy pixels may still be present on
the detector shadowgram. They may not
be visible on an individual detector image, but can be revealed by
those accumulated over several observations.

To estimate the effect induced by ``noisy'' pixels, we simulated a
detector image for a typical \textit{ScW} exposure of $2$~ks, and
inserted one pixel in an arbitrary position exceeding the mean
detector count rate ($40$~cnts/pix/\textit{ScW}) by a factor of
$\sim2.5$ ($\sim9.5$ standard deviations). For a typical \textit{ScW}
such a weak pecular pixel introduced negligible systematic noise and
the reconstructed sky was dominated by Poisson statistics. However,
with increasing exposure, the effect became more significant. We
accumulated the mosaic image of a $280$~ks staring observation of
NGC~4151 with a simulated detector and one hot pixel. When the
position of a noisy pixel was randomly distributed on a detector in
every staring observation, the total mosaic did not contain any
significant systematic residuals. When a noisy pixel was fixed on the
detector, the standard deviation of reconstructed sky was $\sim20\%$
higher, than without a hot pixel on the detector.

To perform additional ISGRI pixel cleaning, we followed the general
approach also employed by \cite{eckert2008}. We stacked detector
images obtained during the spacecraft orbit after removing flux from
known X-ray sources (Sect.~\ref{section:iros}), detector background
map $B$ (Sect.~\ref{section:detector}), and Galactic X-ray background
(Sect.\ref{section:ridge}). The distribution of pixels on a stacked
detector image was well described by Gaussian with zero mean. Thus we
expected that $99.7\%$ of the pixels have a value in the range
$[-3\sigma,+3\sigma]$.  The ISGRI detector contains $128\times128$
pixels which gives us $\sim50$ of them with an expected value greater
than $3\sigma$. However, we typically detected $200-300$ deviations
from zero to larger than $3\sigma$. We removed these pixels from
further analysis.  Filtering done in this way reduces the detector
area by $\sim2\%$, and has a minor effect on flux from serendipitous
faint sources.

\section{Galactic background}
\label{section:ridge}

\begin{figure}
\centerline{
 \includegraphics[width=0.25\textwidth,height=0.24\textwidth]{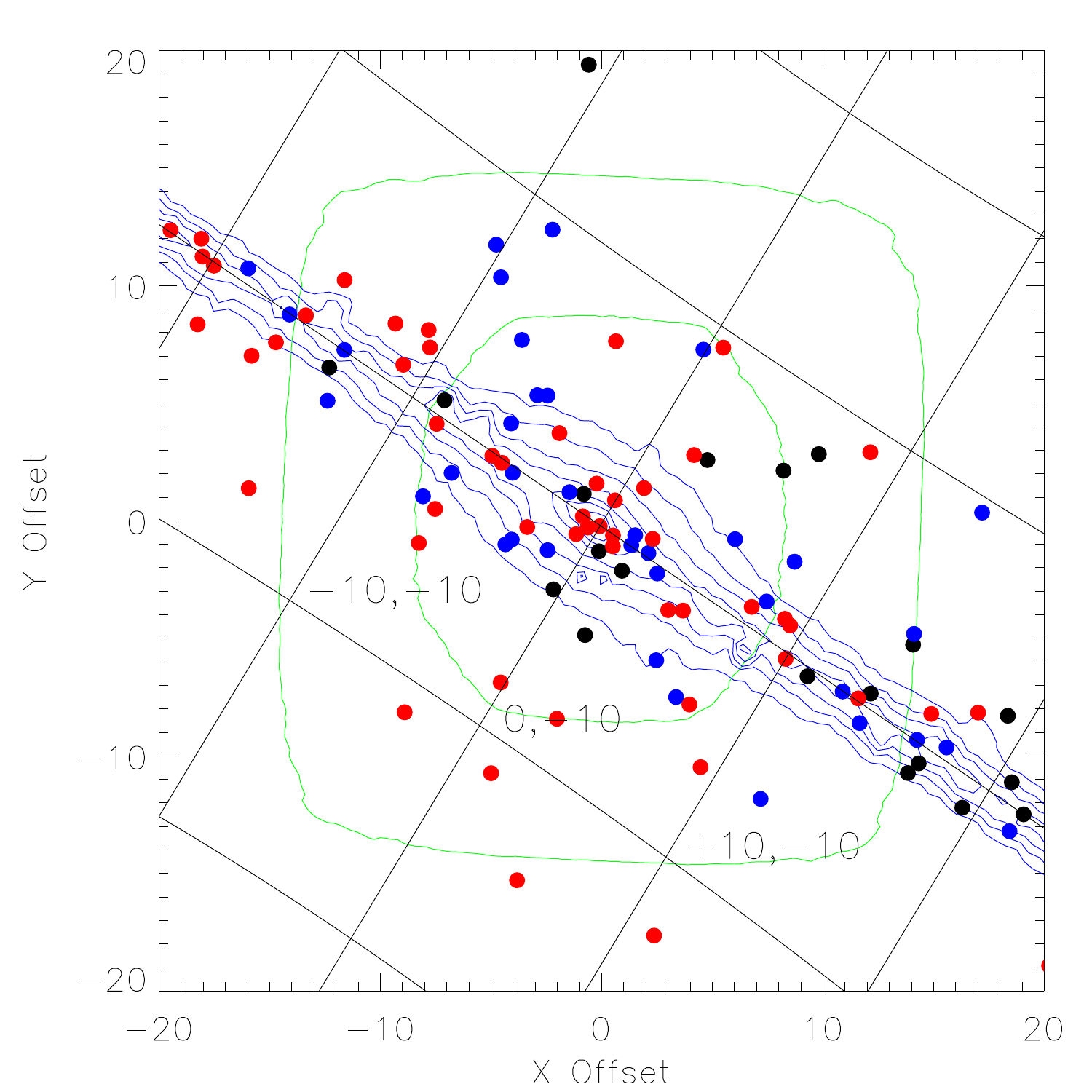}
 \includegraphics[width=0.25\textwidth,height=0.24\textwidth]{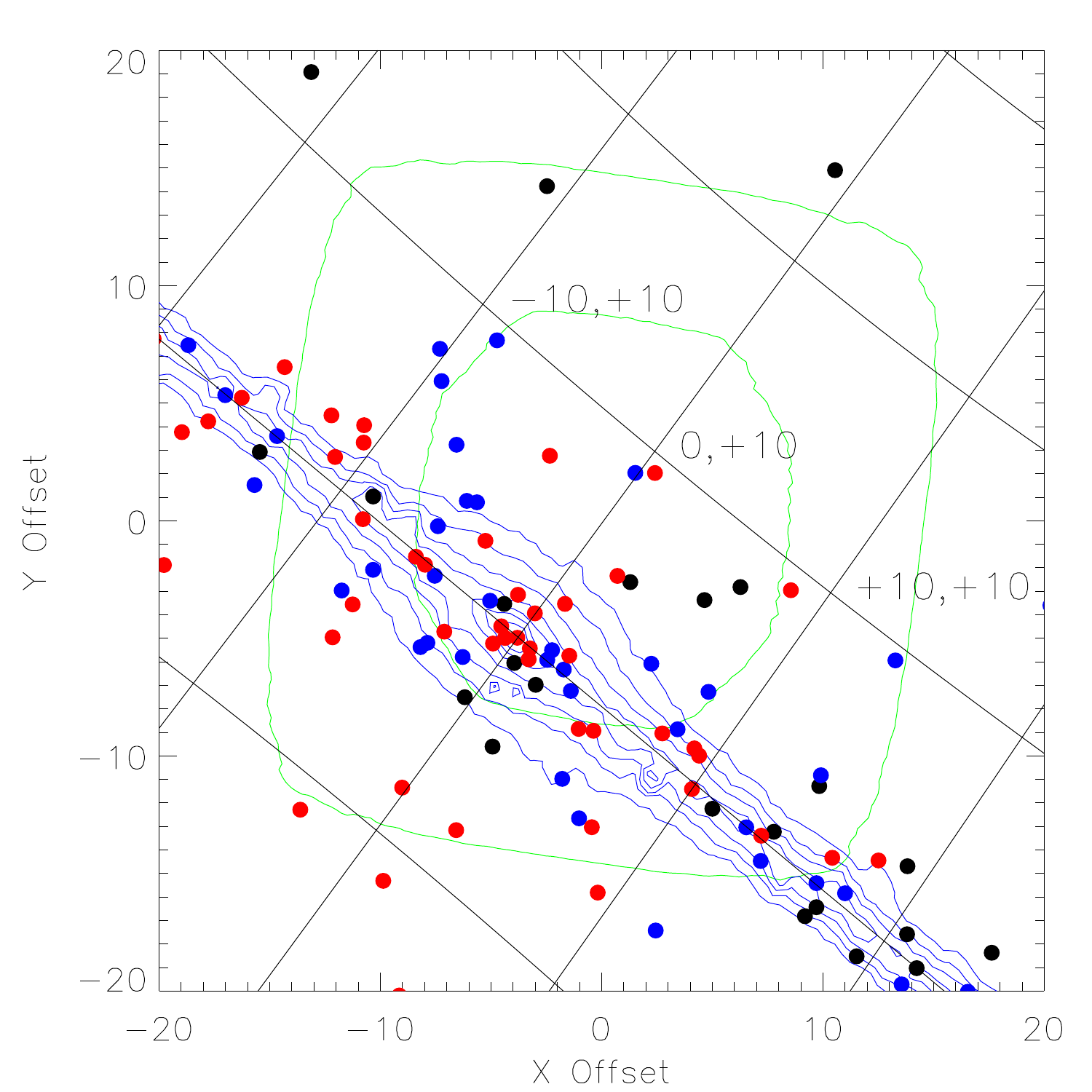}
}

\caption{Two relative alignments of GP in the IBIS FOV, left centered
  and the right shifted along the galactic latitude. The blue contours
  are isophotes of the $4.9\mu m$ surface brightness of the Galaxy
  (COBE/DIRBE) revealing the bulge/disk structure of the inner
  Galaxy. The NIR brightness of the galaxy traces the hard X-ray Ridge
  emission.  The small and large rounded squares on each plot
  demonstrate the full and partial coded areas respectively. The
  points show sky positions of the hard X-ray sources detected on the
  $20$~Ms time-averaged map
  (Fig.~\ref{fig:gplane},\ref{fig:ghosts}). A signal-to-noise ratio of
  the sources is shown by black ($5-10$), blue ($10-30$), and red
  ($>30$). The difference of detector illumination by the Ridge is
  shown in Fig.~\ref{fig:ridcnts}.}\label{fig:skyrid}
\end{figure}

During observations of the Galactic Center region the IBIS field of
view contains many discrete sources (Fig.~\ref{fig:skyrid}). But in
contrast to high galactic latitude observations, the underlying sky
background is not flat. From early X-ray observations we know, that
the Galaxy reveals itself as a strong diffuse emitter
\cite[e.g.][]{worrall82}.  The morphology of the Galactic X-ray
background at energies above 20 keV is now relatively well known. As
shown in recent RXTE and INTEGRAL investigations
\citep{revetal06b,krietal07a}, the X-ray background is traced by the
near infrared brightness of the Galaxy (blue contours in
Fig.~\ref{fig:skyrid}).  We will refer to the Galactic X-Ray
Background later as ``Ridge'' emission or GRXE (Galactic Ridge X-ray
Emission).

The measured 17-60 keV GRXE intensity per IBIS FOV reaches 200 mCrab
in the region of the Galactic bulge. Such strong emission will not
appear on a deconvolved IBIS/ISGRI sky, because the coded mask
technique is not suitable for building an image of sources with
angular sizes larger than the telescope{'}s angular resolution
(12$^\prime$). However, the presence of Galactic X-ray Background in
the IBIS FOV strongly affects the overall shape of the ISGRI detector
shadowgram, which leads to appearence of systematic noise on the
reconstructed sky images.

\begin{figure}\centerline{
 \includegraphics[width=0.25\textwidth]{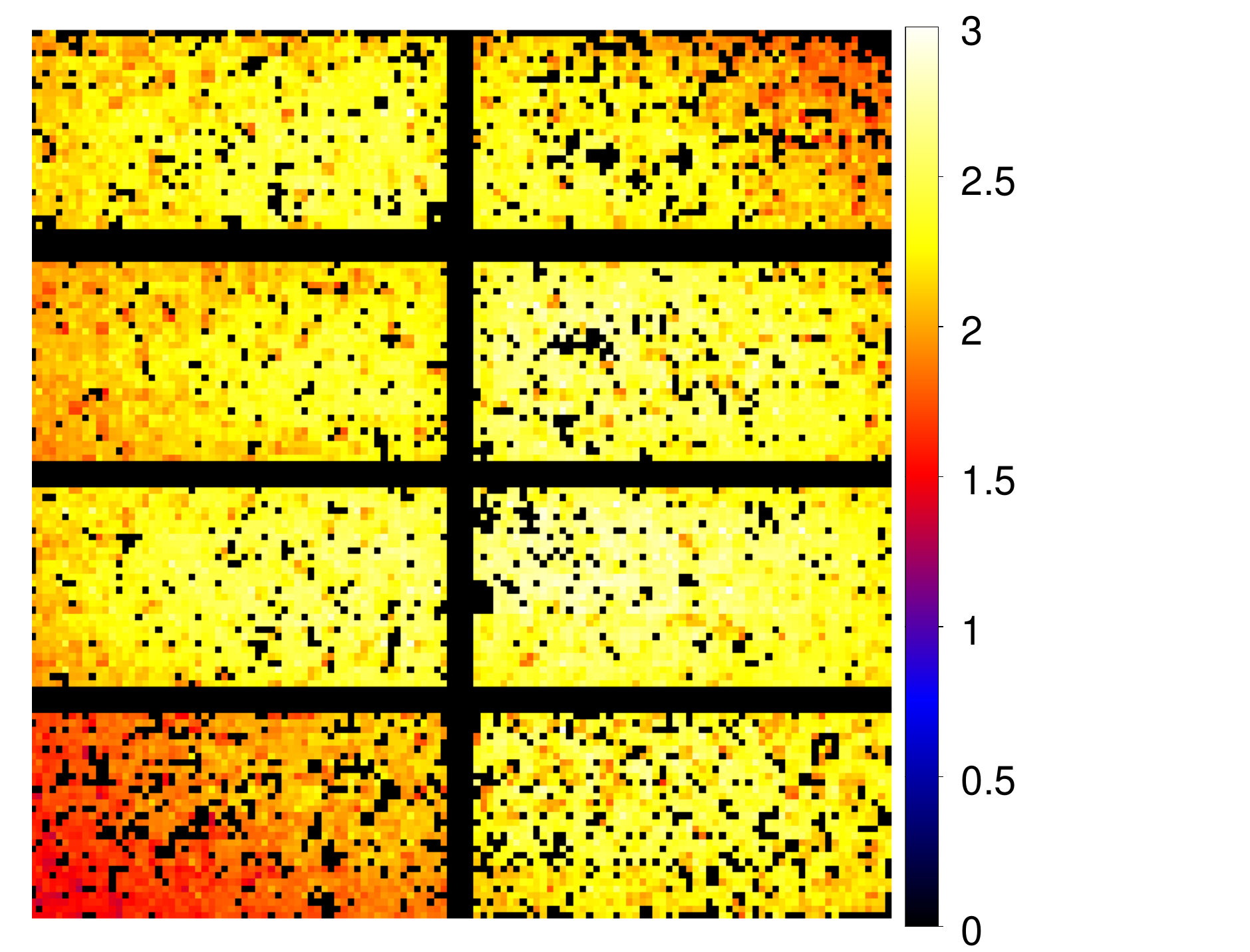}
 \includegraphics[width=0.25\textwidth]{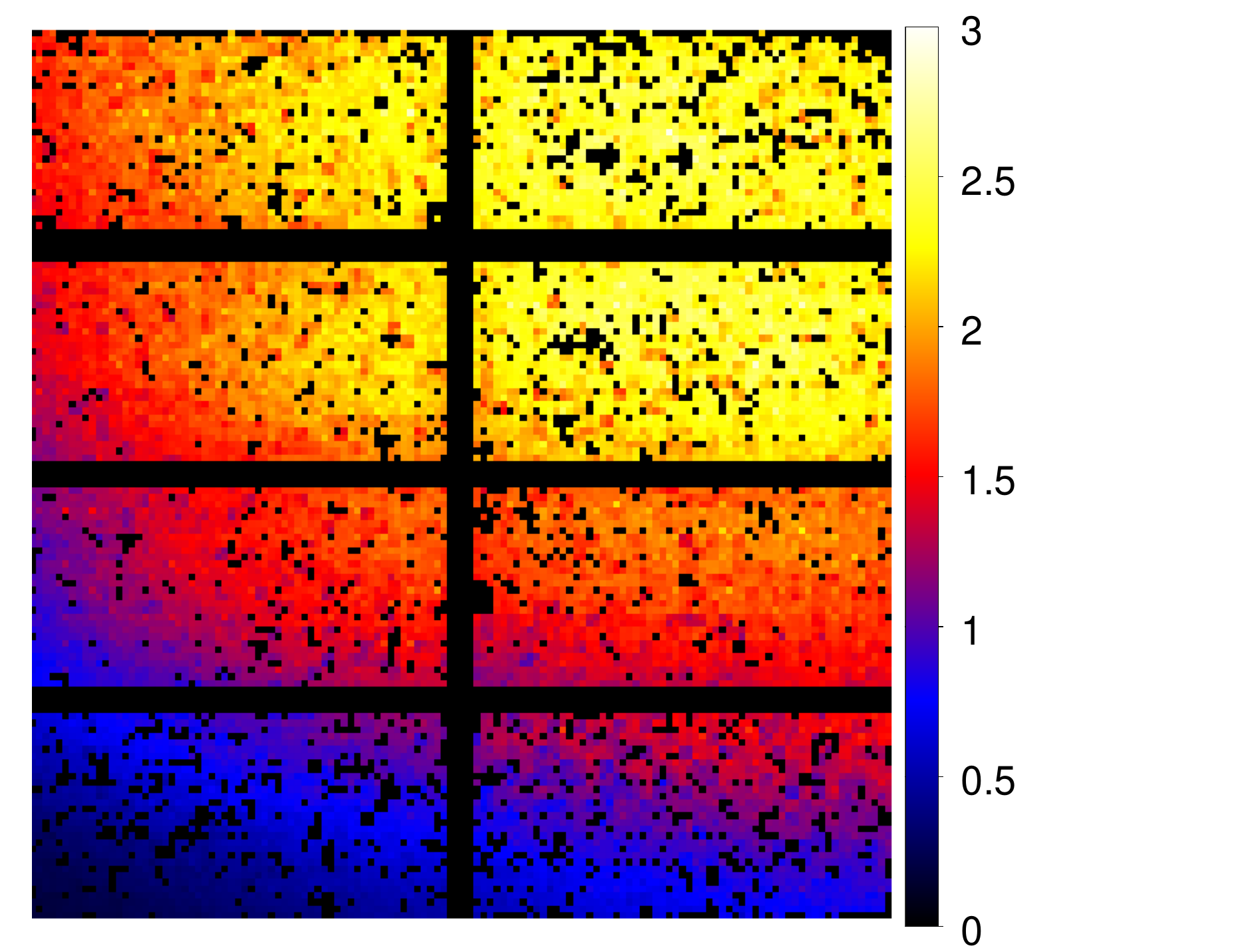}}

\caption{The contribution of Galactic X-ray Background to the ISGRI
  detector count rate for different alignments of GP in the IBIS FOV
  (see also Fig.~\ref{fig:skyrid}). The detector count rate
  normalization is estimated from actual observation of GC in the
  energy range 17-60~keV. The corresponding images in
  Fig.~\ref{fig:skyrid} and \ref{fig:ridcnts} are co-aligned in
  detector coordinates. The Ridge, appeared in the given FOV corner,
  illuminates the opposite corner in the detector
  image.}\label{fig:ridcnts}
\end{figure}

The detector exposure by the point source and GRXE is different. The point
source forms a shadow on the detector. Each exposed pixel contains an
approximately constant source count rate, therefore the source
shadowgram is flat (in illuminated pixels). This is explicitly assumed
in Eq.~\ref{eq:detcnts1}. When IBIS FOV contains the Galactic Ridge
emission, one detector pixel ``sees'' different parts of the Ridge
through the mask{'}s open elements. The resulting detector image is convolution of the
Ridge sky brightness distribution with the IBIS mask and the
collimator response.

In order to model the contribution of the Ridge component to the
detector we have to assume some predefined sky surface brightness
distribution. The Ridge emission at energies $17-60$~keV is well
traced by the near infrared brightness of the Galaxy. The volume
density distribution has been intensively investigated by COBE/DIRBE
\citep[see e.g.][]{dwek95}. We use the Galactic disk and bulge model
from \cite{revetal06b}, which describes the COBE/DIRBE data
in the simplest way. The model was taken from the near-infrared data
and renormalized to fit X-ray intensity measured by INTEGRAL. We
produced sky brightness maps by integrating the model volume density
through the line of sight.

The contribution of the Ridge emission to the ISGRI detector can be
implemented with the so called ``gray mask'' approach: the given
detector pixel sees the Ridge emission through the open mask elements
in a solid angle limited by the telescope collimator.

To demonstrate Ridge detector exposure, we projected GRXE sky
brightness onto the detector using two positions relative to the IBIS
FOV. In position ``A'' the central part of the Ridge emission is
placed in FOV on-axis (spacecraft pointing $l=0^{\circ}, b=0$,
Fig.~\ref{fig:skyrid}, left), and position ``B'' when the Ridge is
$7^{\circ}$ away from the FOV center ($l=0^{\circ}, b=7^{\circ}$,
Fig.~\ref{fig:skyrid}, right). The Ridge detector illumination
implemented with the ``gray mask'' is shown in Fig.\ref{fig:ridcnts}.

As clearly seen from the modeled detector images, the Ridge emission
introduces large scale variations on the ISGRI detector. When IBIS is
centered on the Galactic plane (position ``A''), the Ridge
contribution to the detector image has a relatively flat shape with
some curvature. In contrast detector variations have a significant
gradient when the GP intersects FOV several degrees away from its
center (position ``B''). That configuration strongly warps the
detector image, which obviously will affect the source flux estimates
and, consequently, the accuracy of the source shadowgrams removal.

\begin{figure}\centerline{
 \includegraphics[width=0.245\textwidth]{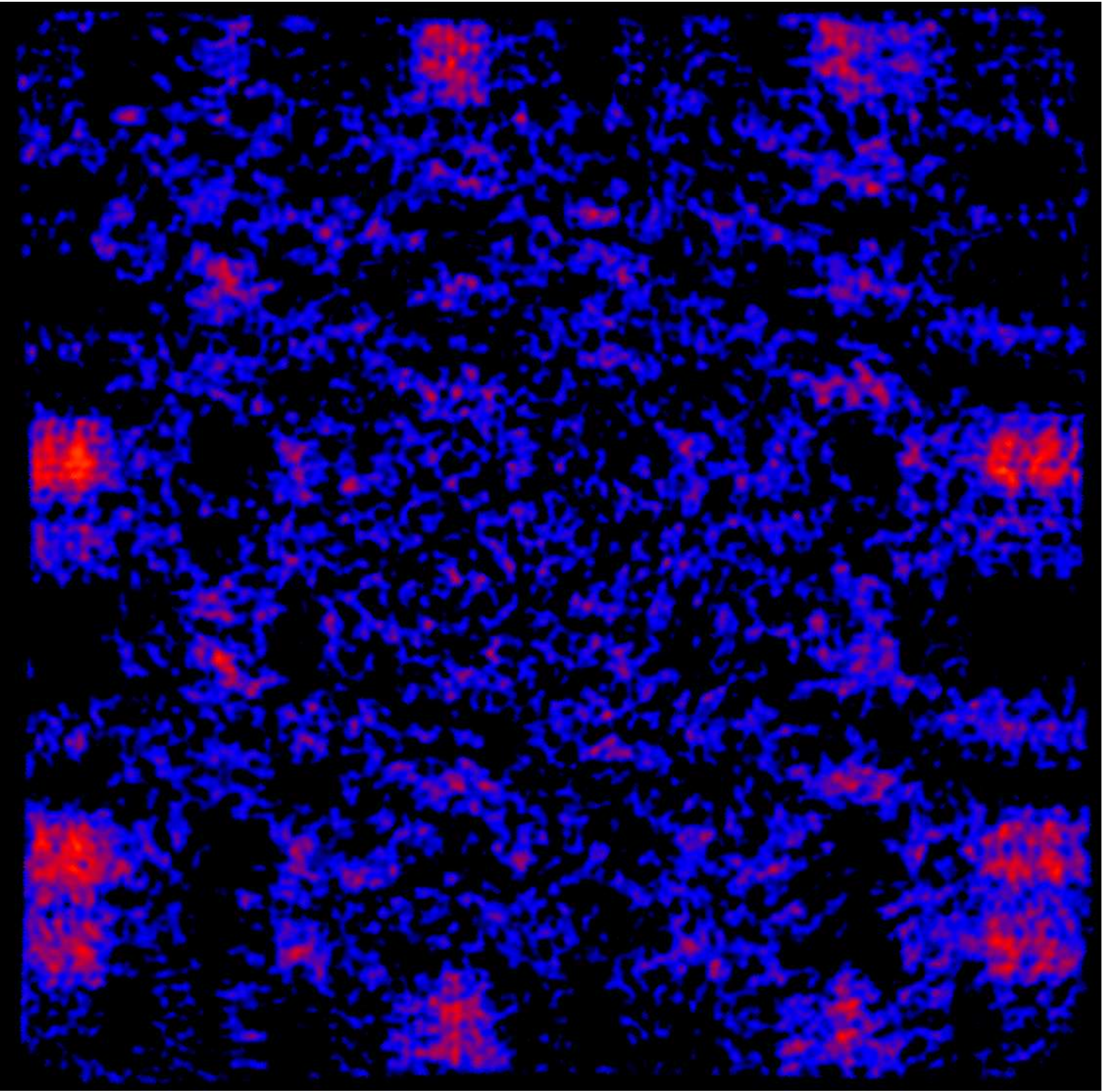}
 \includegraphics[width=0.25\textwidth]{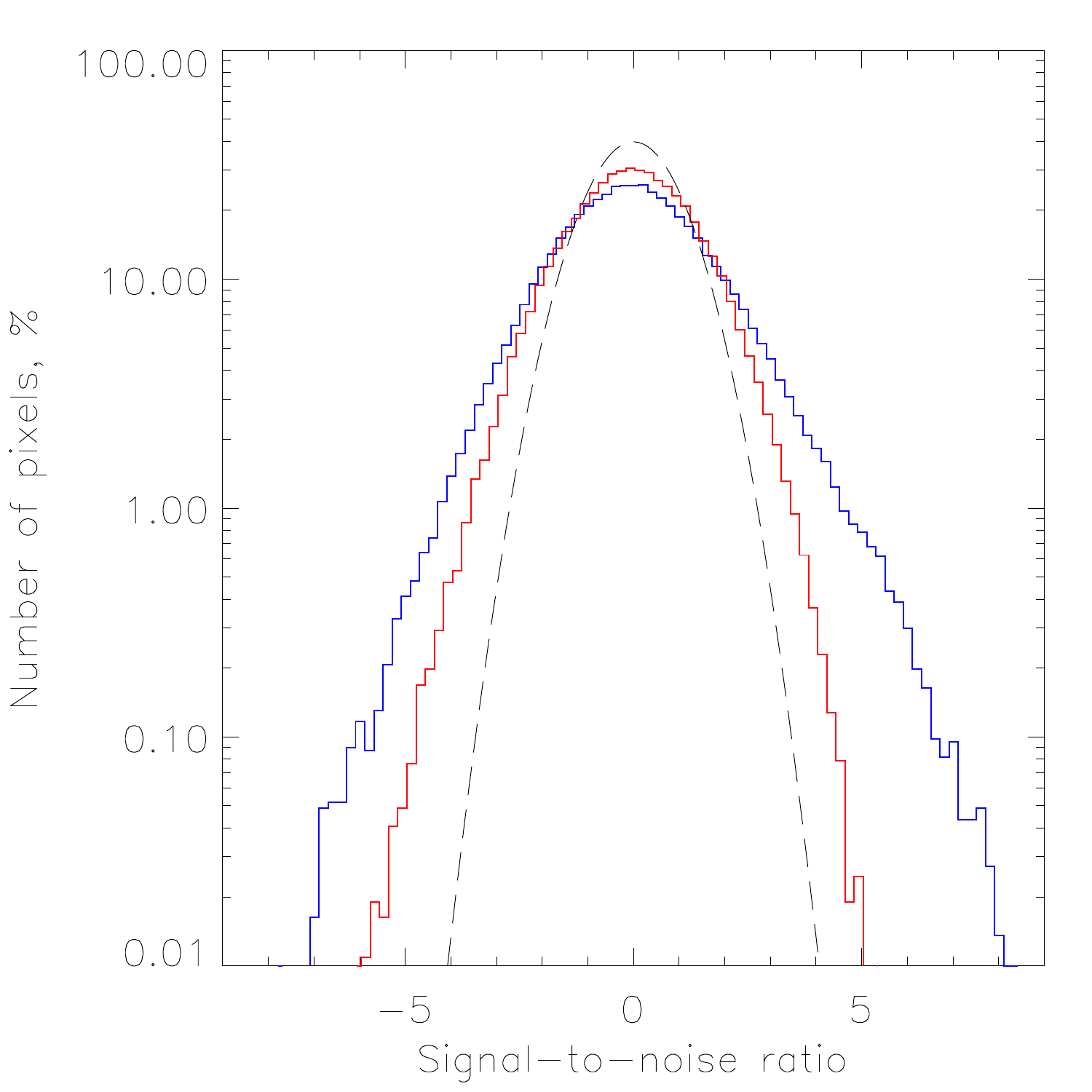}}
 
\caption{\textit{Left:} The sky mosaic of 280~ks staring
  observation of NGC~4151. The Ridge contribution
  (Fig.~\ref{fig:ridcnts}, right) was artificially added to the
  detector plane for each individual observation. For reference
  see the left image in Fig.~\ref{fig:ngc4151}. \textit{Right:}
  Signal-to-noise ratio distribution of a number of pixels in a
  reference and current sky images shown by red and blue
  histograms, respectively. The long dashed line represents the normal
  distribution with unit variance and zero
  mean. }\label{fig:ridcnts:ngc4151}
\end{figure}

The impact of the presence of the Ridge emission in the IBIS FOV to
the reconstructed sky image is significant. The comprehensive
demonstration of this effect directly on the GC data is not possible
due to more serious effects related to the very complicated detector
exposure to many bright sources. That is why, in order to show the
Ridge contribution in the reconstructed sky, we used the relatively
simple and clean 280~ks observation of NGC~4151 for reference
(Fig.~\ref{fig:ngc4151}, left). The artificial detector Ridge
component with actual normalization measured in the GC (position
``B'') was added to the detector image of every spacecraft
observation. Since we used staring observations, this operation was
equivalent to placing Ridge on the sky $7^{\circ}$ away from the
NGC~4151. The quality of the final mosaic
(Fig.~\ref{fig:ridcnts:ngc4151}, left) was very poor and the
signal-to-noise distribution of pixels had strong non-Gaussian wings
(blue histogram in Fig.~\ref{fig:ridcnts:ngc4151}). The standard
deviation of image was $1.7$, in contrast to $1.3$ of the referenced
sky. Thus, we can conclude that the Ridge emission can introduce
significant systematic noise to the reconstructed sky.  However degree
of image worsening depends on exposure and pattern of the
observations.

\begin{figure}\centerline{
 \includegraphics[width=0.5\textwidth]{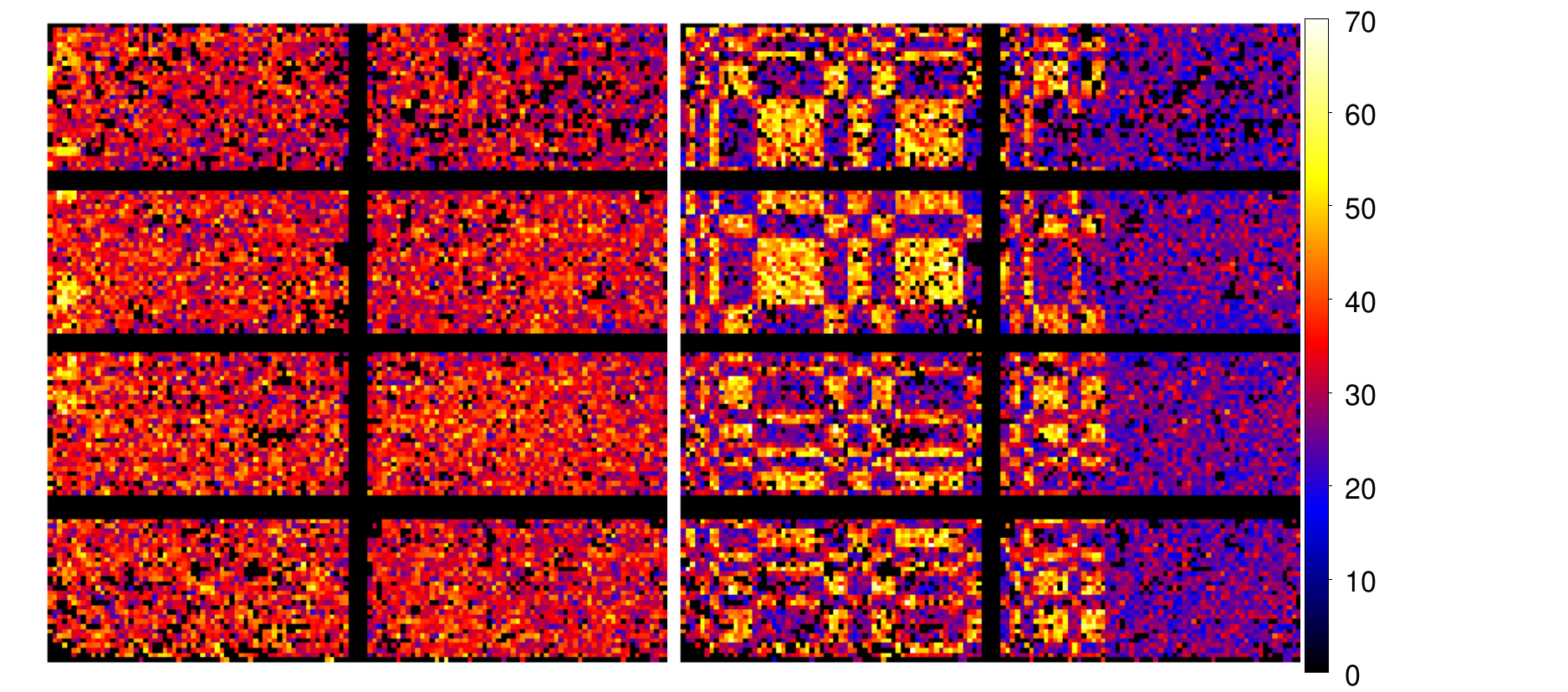}}
\caption{\textit{Left:} The ISGRI detector image of the Galactic Center
  ($l=3.9^{\circ}, b=2.1^{\circ}$). \textit{Right:} The shadowgram of a
  sky region containing only one bright source Sco X-1
  ($l=4.5^{\circ}, b=30.0^{\circ}$).  The observations were separated
  by a 10 hour time interval when the background count rate had not
  significantly changed, i.e. the background map for both images is
  the same. In the right image the background and source counts are
  clearly distinguishable. This leads to the straightforward
  application of the standard coded mask reconstruction algorithm (see
  Sect.~\ref{section:method}). The left image contains many overlaping
  shadowgrams of bright sources. The assumed background model is not
  accurate and the general method is confused. (see
  text)}\label{fig:detcnts}
\end{figure}

Let us consider the typical IBIS/ISGRI observation of the Galactic
Center region (position ``A''), as the most representative observation
containing many bright sources and strong Ridge emission in the
FOV. The detector image is shown in the left frame of
Fig.~\ref{fig:detcnts}.

Firstly, the catalogue of predefined source positions provides about a
hundred objects in the FOV. Among them, $6-8$ bright sources are
usually detected with $S/N>5$ in the individual observation of
$\sim$2.5~ks exposure time.  By summing up all $PIF\textrm{s}$
of these sources, one can show that practically all detector pixels
are illuminated by at least one source (see Fig.~\ref{fig:detcnts},
left). This means, that \textit{any} measurements of source fluxes in
the GC will be affected by the correlations of source shadowgrams,
especially for those having replicated patterns due to the periodic
mask elements. Furthemore, there are no source-free pixels to estimate
the background count rate. Flux measurement, based on the balance
matrix \citep{krietal05,goldwurm03} is not accurate, because the
background map for a given source is strongly affected by other
sources. The general reconstruction algorithm works ``as is'', trying
to estimate the fluxes of the brightest source and leaving many
systematic artifacts on the detector, and consequently, on the
resulting sky image.

The situation is further complicated by the fact, that practically all
ISGRI pixels are illuminated at the same time by the Ridge emission
(see Fig.~\ref{fig:ridcnts}). This leads to a high correlation between
shadowgrams of a manifold of sources and the Ridge.

Summing up all mentioned effects, we conclude, that
Eq.~\ref{eq:detcnts1} based on simple background map $B$ is not
suitable for Galactic observations, and the iterative source removal
is not valid when the detector map is warped by the Ridge. Generally,
sky reconstruction is highly complicated in the case of Galactic
Center observations. To a large extent, the INTEGRAL/IBIS/ISGRI survey
sensitivity is limited by systematic uncertainties, and new exposures
in the GC provide only minor effects on the total sensitivity.

\section{Modified sky reconstruction method}
\label{section:method2}

The sky reconstruction method for IBIS/ISGRI should be able to split
a detector image into the following layers:
\begin{itemize}
\item source $PIF\textrm{s}$. Characterized by a flat count
  rate and the pattern of the projected mask.
\item illumination by the Ridge. This layer
  has certain low-(spatial)-frequency variations. The characteristic
  variations depend on the telescope orientation relative to the
  Galactic Plane and Center.
\item ISGRI background layer. This contains the CXB flux and instrumental
  background, and has a smooth count rate.
\end{itemize}
all these layers correlate in different degrees with each other due to the
shared detector area. We modify Eq.~\ref{eq:detcnts1} to accommodate
the Ridge component $R$:

\begin{equation}
D =  \sum_{i=0}^{M} f_iPIF_i + k_{B}B + k_{R}R
\label{eq:detcnts2}
\end{equation}

The direct simultaneous solution of Eq.~\ref{eq:detcnts2} is
practically impossible before the components are orthogonalized.  At
first, we tried to reduce the mutual correlation between layers using
different spatial features of the layers and the short list of sources
(to do this, we selected $M_b$ bright sources with $S/N>7$ which
appeared in the FOV). We fit the relation (\ref{eq:detcnts2}) with the
modelled Ridge contribution $R$, source \textit{PIF}s, and background
map $B$. The fitting procedure was unstable, producing inadequate
estimates of the layer normalizations. We noticed high uncertainty in
determination of Ridge and Background map components, respectively
$k_R$ and $k_B$, due to its high mutual correlation.

One possible way to further disentangle the Ridge layer from others is
to use different spacecraft orientations relative to GP. To do this,
we attempted to simultaneously fit the nearest $Q$ observations
({\textit{ScW}s) before and after the current one:

\begin{equation}
D_j  =  \sum_{i=0}^{M_b} f_iPIF_{i,j} + k_{B}B_j + k_{R}R_j,
\label{eq:detcnts3}
\end{equation}
where index $j$ is running in the range $[-Q,+Q]$ relative to the
current observation $j=0$. $M_b$ denotes the number of bright sources
which appeared in the FOV of every $j\in[-Q,+Q]$ observation with
$S/N>7$. The on-axis direction of all selected observations is
requested to be inside $15^{\circ}$ radius around the position of
current observation.

We fitted large data sets using Eq.~\ref{eq:detcnts3} for different
$Q>0$ and found the Ridge component $k_R$ much better constrained with
respect to the other layers. Guided by the stability of the fitting
procedure, we chose $Q=6$ as an reasonable selection
criteria. Generally, the procedure is more stable for more spatially
scattered observations in the direction perpendicular to the GP. Among
the available INTEGRAL observational patterns we have found that the
best pattern for this approach is the \textit{Galactic Center Deep
  Exposure} \citep[GCDE, Core Program, see e.g.][]{core} and the
\textit{Galactic Latitude Scans} (PI Sunyaev). The fitting procedure
running on large data sets of the usual 5$\times$5 pattern (see ISOC
Newsletter \#12, September 2004) and the \textit{Galactic Plane Scans}
\citep[GPS, Core Program, see e.g.][]{delsanto03,rodriguez03} is less
stable, but still provides valuable results.

We should note, that the contribution of the $i\textrm{th}$ source
($f_iPIF_{i,j}$) in Eq.~\ref{eq:detcnts3} is estimated under the
assumption of the constant flux $f_i$ during the considered time
interval. In a similar way, the detector background count rate $k_B$
is also considered constant. For the chosen value of $Q=6$, the
maximum number of observations is $13$, which is in total $\sim
25$~ks. Generally, for such a time interval, the majority of
galactic X-ray sources do not vary by a factor of more than $\sim$2,
except during the outburst activity.

Actually, the modified sky reconstruction algorithm described here is
used only for constraining the Ridge component and its following
subtraction. After this step, the IROS procedure
(Sect.~\ref{section:iros}) is applied in the usual manner on the
detector plane of the given observation (ScW).

Note, that the employed procedure still doesn{'}t allow us to
completely resolve the problem of highly correlated shadowgrams on the
detector in the case of the Galactic Center observations, but it at
least reduces the systematic residuals on mosaic images introduced by
Ridge emission.

\subsection{Removing systematic residuals from sky images}
\label{section:atrous}

After removing the source shadowgrams and the background from the
detector we still see the systematic effects on sky images. These
residuals can be clearly seen on deep extragalactic observations in so
called staring mode. In these observations orientation and the roll
angle of the telescope is fixed. The sky images are stacked
(pixel-to-pixel) in detector coordinates. All systematic residuals,
not visible on images of individual observations (with an exposure of
$\sim$2~ks) are amplified on the stacked image. The sky region around
NGC~4151, was reconstructed with the help of the general deconvolution
algorithm from the data collected in the staring mode, as shown in the
left image of Fig.~\ref{fig:ngc4151}. The characteristic
chessboard-like squares and ripples are clearly seen in the fully
coded (central $10^\circ\times10^\circ$ square) and partically coded
(outer parts) field of view.

Due to the absence of bright sources in the field of view during these
staring observations, it is clear that the presence of the patterns
seen in Fig.~\ref{fig:ngc4151} does not depend on the accuracy of our
model of source shadowgram. We cannot also attribute visible
systematic artifacts to the detector noisy pixels, because they were
filtered out (Sect.~\ref{section:pixels}).

The major origin of these patterns on the sky is the limited knowledge
of the background pattern on the detector. It may consist of several
unaccounted for parts, like the unexpected variations of pixel gains,
the effective lifetimes, and lifetimes of the detector modules (the ISGRI
detector has 8 modules of pixels, which in many cases change, see
e.g. Fig.~\ref{fig:ridcnts}). Judging from a particular pattern of
residuals on the sky, the major effect is due to the inaccurate estimate
of efficiency and the energy band-passes of the detector pixels.

In the context of the general sky reconstruction method, it becomes
extremely difficult to correct this problem due to the continuous degradation
of detector pixels and the variation of the background environment. Therefore,
while understanding the origin of the residual structures on the sky
we decided to implement an alternate solution to this problem.

As the spatial scale of systematic artifacts on the sky is
significantly different from that of the point sources, we have
implemented the wavelet-based image filtering procedure.

The key point of all wavelet methods is that the wavelet transform
(WT) is able to discriminate structures as a function of the spatial
scale, and thus is well suited to detect small scale structures on an
image embedded within larger scale features. That is why, WT has been
widely used for the structure analysis of galaxy clusters
\citep{slezak94,grebenev95,rosati95,biviano96,vikhlinin96}.

In the context of an individual IBIS/ISGRI observation, we are
interested in removing all large scale structures from non-uniform sky
background rather than in the detection of point sources. The task is
greatly simplified by the fact, that the coded-mask aperture technique
is not able to reconstruct an image of objects with a spatial size
greater than the angular resolution of the telescope. For point source
detection all structures more extended than a point source can be
safely removed. In other words, we do not need any thresholds to
discriminate noise and signal, we can remove systematic residuals with
a given angular scale ``as is''. In this way, the WT works as a
non-parametric sky background approximation. In the similar way, the
wavelet transform was used for subtraction of nonuniform background
and for filtering images obtained with the coded-mask X-ray telescope
ART-P aboard the GRANAT mission \citep{grebenev95wv}.

\subsection{Wavelet decomposition method}
\label{section:atrous:method}

To decompose a sky image, we use the \textit{\`a~trous}\/ digital
wavelet transform (DWT) algorithm because it allows easy
reconstruction \citep{starck94,slezak94,vikhlinin96}. The method uses
a kernel $K_J = F_{J} - F_{J+1}$, where integer $J$ is the so called
spatial scale index. Each $F_J$ is constructed by five weighting
coefficients $[1,4,6,4,1]/16$ spaced by a $2^{J-1}$ interval. Note,
that each $F_J$ can be roughly approximated by a Gaussian of width
$2^{J-1}$ ($F_1$ is a $\delta$-function). The convolution of an image
with $F_J$ preserves flux ($\Sigma F_J=1,\forall _J$), and convolution
with $K_J$ emphasizes the structures with the characteristic size
$\approx 2^{J-1}$~pixels, or $2^{J+1}$~arcminutes in the case of the
IBIS/ISGRI image ($1\textrm{pix}\approx4\arcmin$). Thus, low $J$
values correspond to small spatial variations or high frequency, and
high $J$ reflects large spatial variations or low frequency.

On the largest scale, $N$ the kernel $K_N=F_N$. The original image $I$
can be easily decomposed to its convolutions, $W_J$ (``wavelet planes of
scale $J$''), with kernels $K_J$:
\begin{equation}
  \label{eq:wv:restore}
  I=\Sigma_{J=1}^N W_J.
\end{equation}
Therefore, we can consider $W_J$ an image containing ``flux'' on
scale $J$; the sum of all fluxes yields the original image. This is the
basis of the wavelet decomposition algorithm. The original image is
convolved with the wavelet of the first scale. The wavelet plane on the
first scale is removed from the image. We then go to
the next scale $J+1$. In other words, we remove all small-scale
features from the image before working at larger scales. That is why
the small scale features (high spatial frequency) do not affect the
convolution at larger scales (at low spatial frequency). Obviously, this
algorithm greatly reduces the interference of the point source with large-scale kernels.


In order to remove the large-scale systematic structures seen in the
left image of Fig.~\ref{fig:ngc4151}, we start with deconvolved
source-free sky image of individual observation: 1) flux image is
decomposed to the wavelet planes. The systematic residuals are
clearly seen at scales $J=5,6,7$. After that, we 2) restore the
original image by Eq.~\ref{eq:wv:restore} omitting these scales, and
3) return the source fluxes to the sky as described in the general
reconstruction method \citep{krietal05,goldwurm03}.

We should stress, that DWT filtering is used on sky flux images,
where systematic residuals are emphasized. The sky variance map was
not filtered, because it contains only formal uncertainties related to
the exposure time for a given sky direction.

\subsection{Impact on point sources}
\label{section:atrous:source}

\begin{table}
\begin{center}
\caption{Reconstruction of the point source flux with different sets of wavelet planes.}
\label{tab:dwtflux}
\begin{tabular}{lcc}
\hline 
\hline 
\noalign{\smallskip}  DWT scales, J & Flux fraction & Significance \\ 
\noalign{\smallskip}\hline\noalign{\smallskip} \hline

$1$ & 0.00435 &     1.54	\\ 
$1,2$ & 0.36121 &   127.95	\\ 
$1-3$ & 0.78648 &   278.60	\\ 
$1-4$ & 0.94208 &   333.72	\\ 
$1-5$ & 0.98549 &   349.09	\\ 
$1-6$ & 0.99647 &   352.98	\\ 
$1-7$ & 0.99913 &   353.92	\\ 
\hline
\multicolumn{3}{l}{original sky image}  \\
$1-8$ & 1.00000 &   354.23	\\ 
\hline
\multicolumn{3}{l}{wavelet scales used in this work}  \\
$1-4,8$ & 0.94210 &   333.72	\\ 

\hline
\noalign{\smallskip}\hline\noalign{\smallskip}
\end{tabular}
\end{center}
\end{table}

According to the above section, the cleaning procedure operates on the
source-free images, i.e. images free from catalogued
sources. Therefore, DWT filtering does not directly affect the flux of
known sources.

However, if position of the source is unknown, its flux will not be
removed from the detector before sky deconvolution, which means that
source will appear on the sky. In this case, the DWT filtering will
clean sky background with embedded point source.

The important question is how the implemented DWT filtering affects
the flux of such an point source. The main idea is that, the DWT
procedure must not significantly change the point source flux.

In order to investigate this issue, we used standard on-axis
observation of the bright source Crab Nebula with total exposure
$2.7$~ks. We performed DWT decomposition of the reconstructed sky
image of Crab, for the range of scales $1-8$.  The image was then
reconstructed by summing up selected wavelet scales. Crab flux was
estimated on the final sky image, convolved with the effective point
spread function \citep{krietal07b}.

The Table~\ref{tab:dwtflux} contains the measured Crab flux for the
different sets of wavelet scales.  The flux is expressed as a fraction
relative to the flux in the original sky image. As seen from the
table, $\sim94\%$ of the point source flux resides in the
high-frequency wavelet scales $1-4$, and the rest contributes to the
low-frequency scales $5-8$.

If we assume that the point source is significantly affected when
$\sim5\%$ of its flux is greater than the $1\sigma$ survey detection
threshold, then wavelet filtering distorts only the point sources at a
detection level of $>20\sigma$. Obviously, such strong known sources
are removed from the detector shadowgram before sky deconvolution, and
returned to the sky image after the DWT filtering in steps 1-2. New
sources detected in the survey, generally have a detection
significance of less than $10\sigma$. We conclude, that the
implemented wavelet filtering method does not affect the flux of point
sources, and does not introduce significant distortion to the survey
sensitivity.

\subsection{Extragalactic sky}
\label{section:atrous:sky}

\begin{figure}\centerline{
 \includegraphics[width=0.25\textwidth]{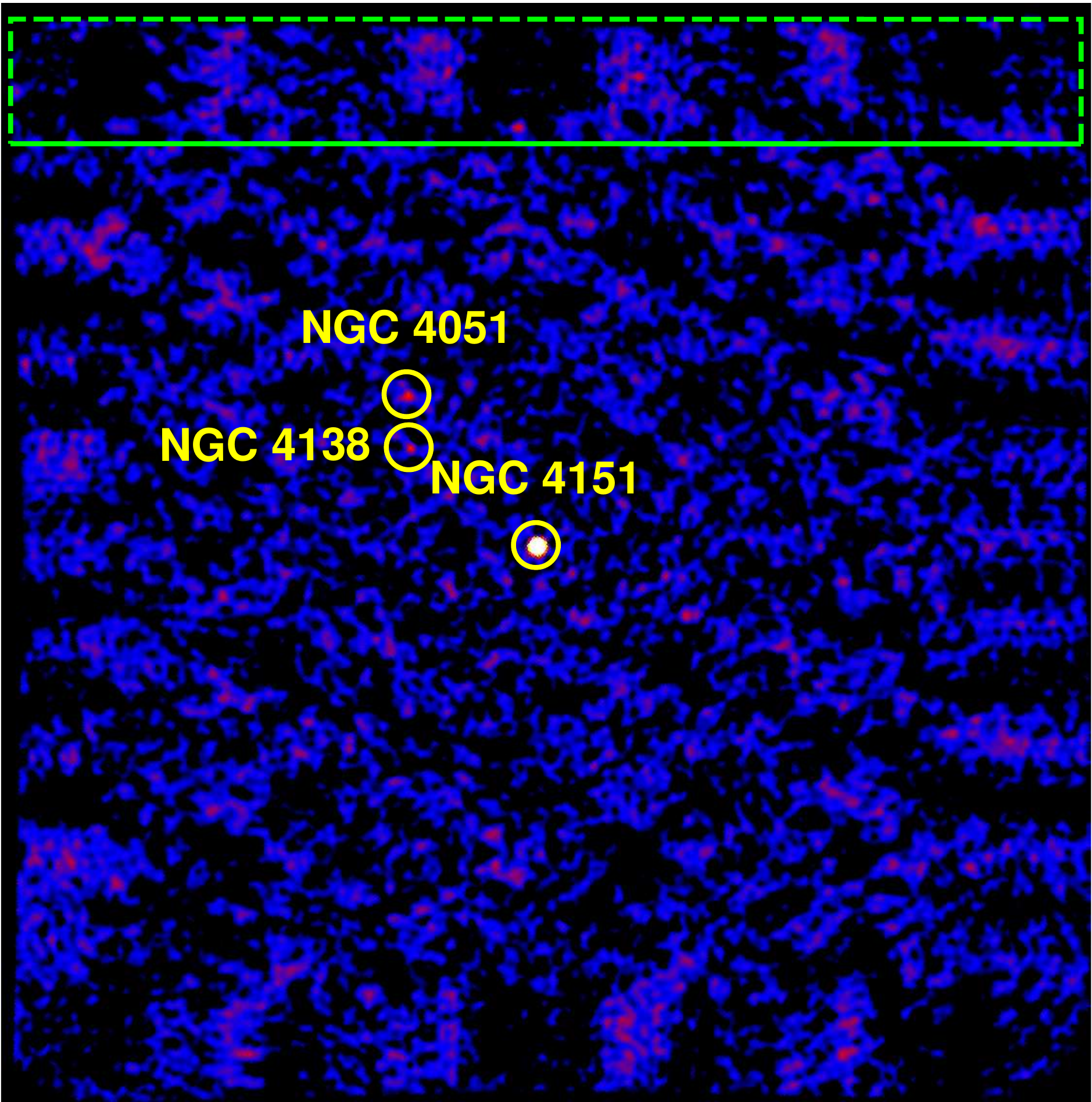}
 \includegraphics[width=0.25\textwidth]{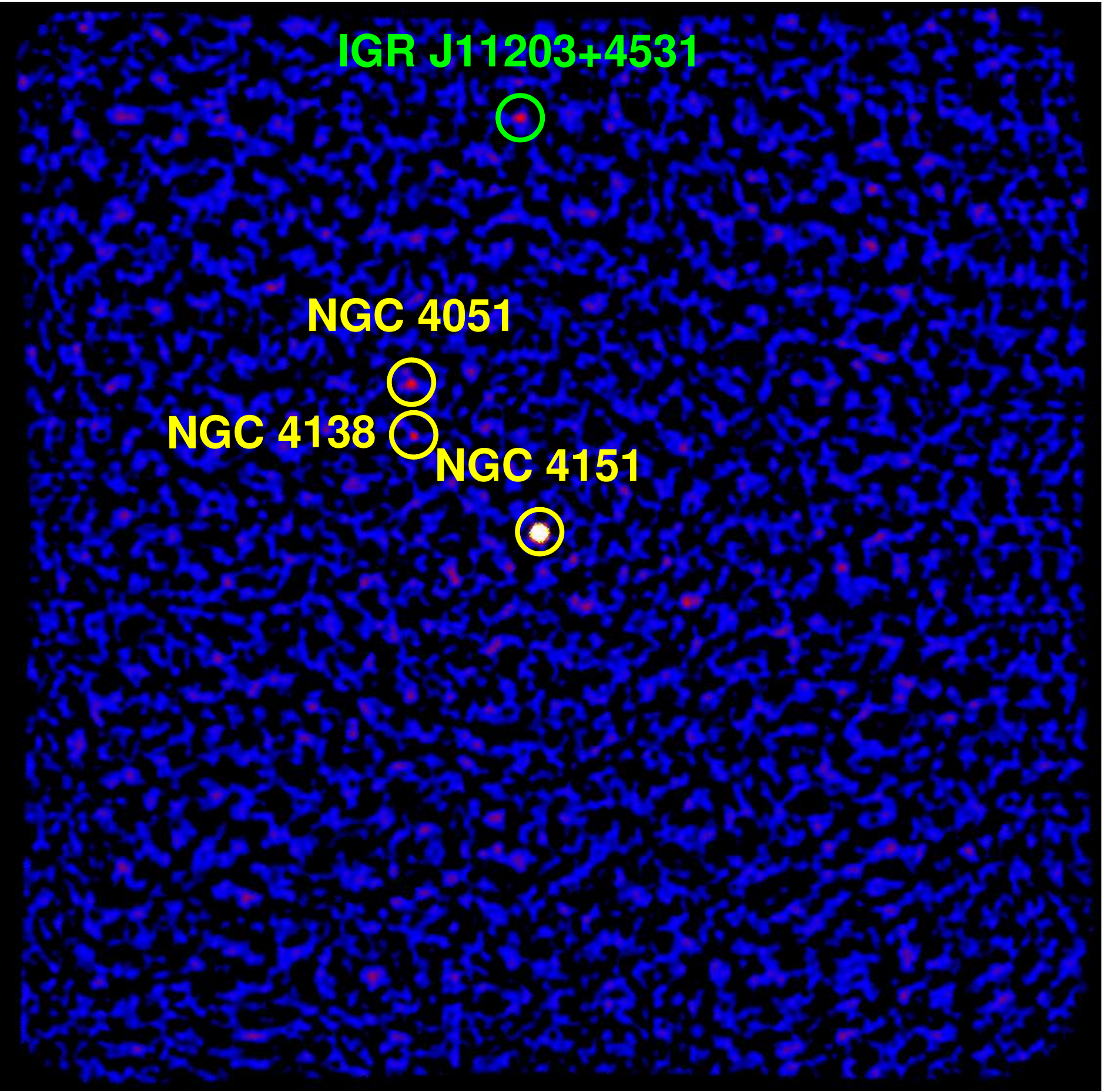}}
\caption{Sky region around NGC~4151 accumulated with a sequence of 95
  staring observations (spacecraft orbits 74--76). The total dead-time
  corrected exposure is 280~ks. The angular size of each image is
  $30^{\circ}\times30^{\circ}$. The left mosaic image is obtained by
  the general method (Sect.~\ref{section:method}). The image on the
  right was produced by summing up images of individual observations
  corrected with \textit{\`a~trous}\/ wavelet decomposition algorithm
  (Sect.~\ref{section:atrous}). The standard deviation of source-free
  pixels on the left and right images relates as 1.3 and 1.0,
  respectively. The new hard X-ray source IGR~J11203+4531, detected
  during this observation is labeled in green. To illustrate how the
  algorithm works, we extract the vertical profile from the green
  rectangular region in the left image
  (Fig.~\ref{fig:ngc4151:demowv}).}\label{fig:ngc4151}
\end{figure}

\begin{figure}
 \includegraphics[width=0.5\textwidth]{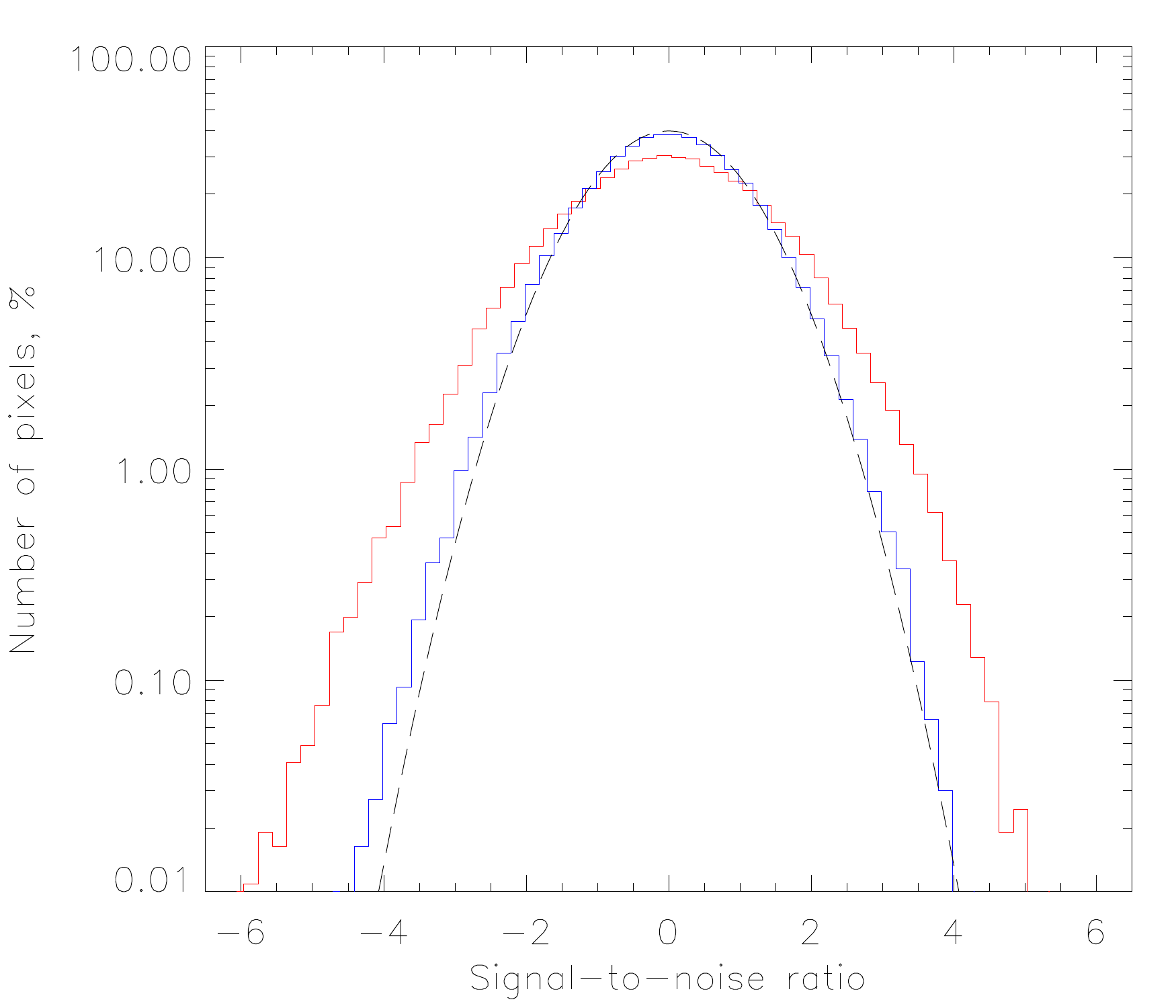}
\caption{Signal-to-noise ratio distribution of a number of pixels in a
  280~ks staring mode deep observation of NGC~4151. The SNR
  distribution of the image obtained with the general method
  (Fig.~\ref{fig:ngc4151}, left) is shown in red. The blue histogram
  shows pixel variance distribution of the sky mosaic accumulated with
  WT cleaned images (Sect.~\ref{section:atrous}). The dashed line
  represents the normal distribution with unit variance and zero
  mean. Assuming normal distribution of $N=183606$ pixels on a
  cleaned image, we expect 5 occurrences at a significance level
  $\sigma=\pm4.2$. However, we detect 12 and 2 excesses at negative
  and positive values, respectively.}\label{fig:sigdistr_ngc4151}
\end{figure}

We performed DWT cleaning procedure (Sect.~\ref{section:atrous:method})
for each observation of the already mentioned NGC~4151 staring mode
campaign. The resulting image is presented in Fig.~\ref{fig:ngc4151},
right.  It is clearly seen that all large-scale artifacts are totally
removed, leaving a clean sky image. In order to demonstrate the
achieved improvement, we built a distribution of signal-to-noise
ratios (SNR) for pixels on the source-free sky image. The SNR
distribution of the original image (Fig.~\ref{fig:ngc4151}, left) is
represented in Fig.~\ref{fig:sigdistr_ngc4151} by the red
histogram. It can be approximated by a Gaussian with
$\sigma\approx1.3$, which is consistent with the measured standard
deviation of image pixels. The SNR histogram of the wavelet filtered
sky image is plotted in blue. The last is well approximated by normal
distribution with a unit variance and zero mean, which means that the
systematic noise has been completely removed. Thus, we conclude that,
for the simplified case of extragalactic observation, the DWT
filtering significantly supresses (practically removes) systematic
noise.

Till now, about 3~Ms of INTEGRAL exposure time has been performed in
staring mode. Usually, these observations are excluded from sky
mosaics due to high systematics. Cleaning these observations with DWT
we can add them to the survey. 

It is interesting to note, that by averaging the archival staring
observation of NGC~4151 we detected a new transient source
IGR~J11203+4531 at sky position R.A.=11h20m21.60s, Decl.=45d31m48.0s
(equinox 2000.0, uncertainty $~4$ arcmin). The source was found at the
FOV edge where strong systematics prevented its detection before (see
Fig.~\ref{fig:ngc4151} and \ref{fig:ngc4151:demowv}).  The source is
seen at $S/N=5.3$ on the original mosaic with $RMS=1.3$, which gives
$4\sigma$ excess. On the cleaned sky the systematics is gone, and the
source has $S/N=5.7$ on the image with $RMS=1.0$, i.e. the source is
revealed at $5.7\sigma$.  This is demonstrated
on the average image profiles shown in
Fig.~\ref{fig:ngc4151:demowv}. The follow-up Swift/XRT observation of
IGR~J11203+4531 revealed two nearby sources with coordinates
R.A.=11h20m26.92s, Decl.=+45:34:53.77 and R.A.=11h20m33.76s, Decl.=
+45:28:17.92 (error radii according to the ``xrtcentroid'' program are
5.96 and 5.31 arcsec, respectively).

\begin{figure*}[t]\centerline{
 \includegraphics[width=0.45\textwidth]{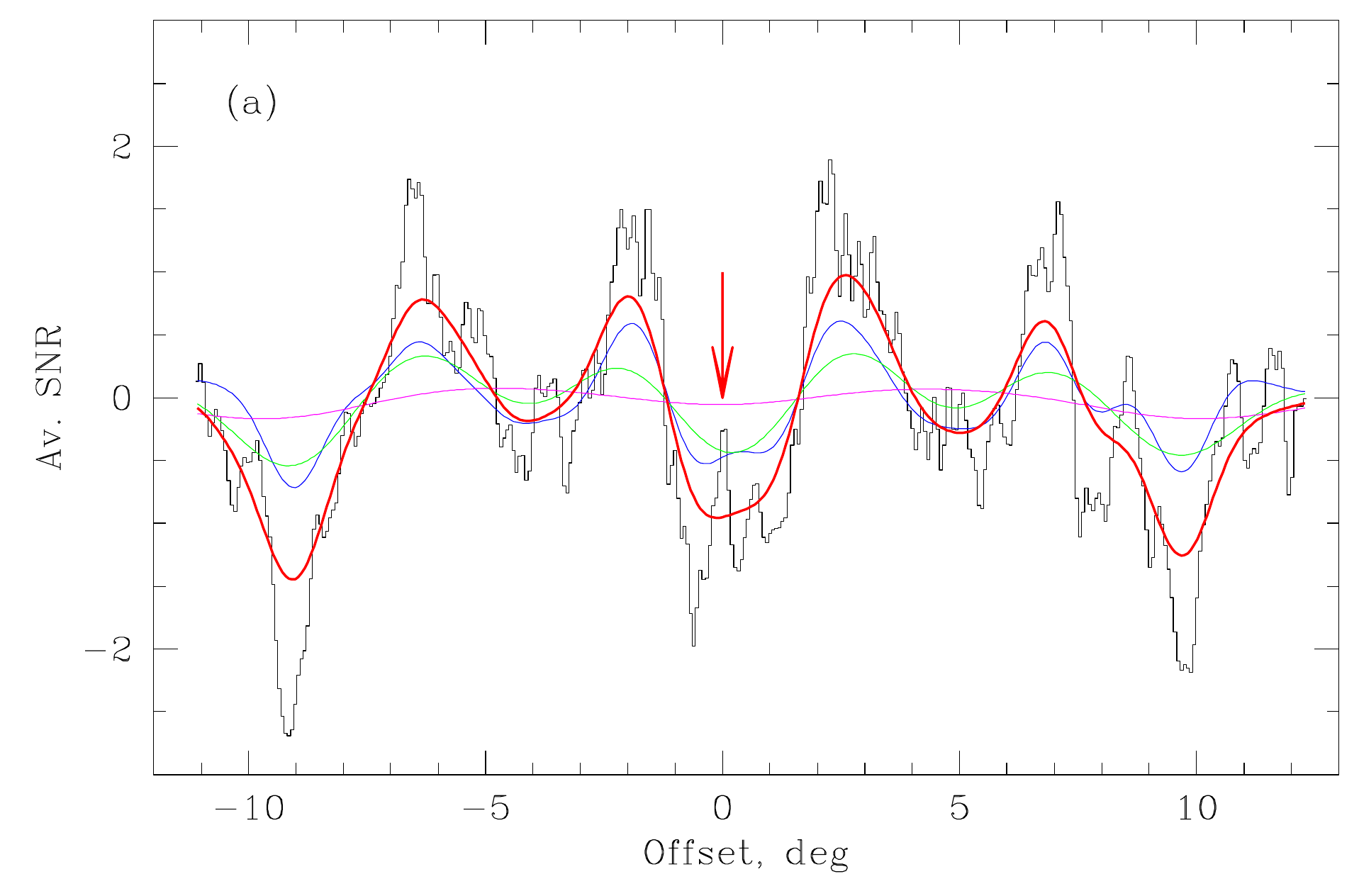}
 \includegraphics[width=0.45\textwidth]{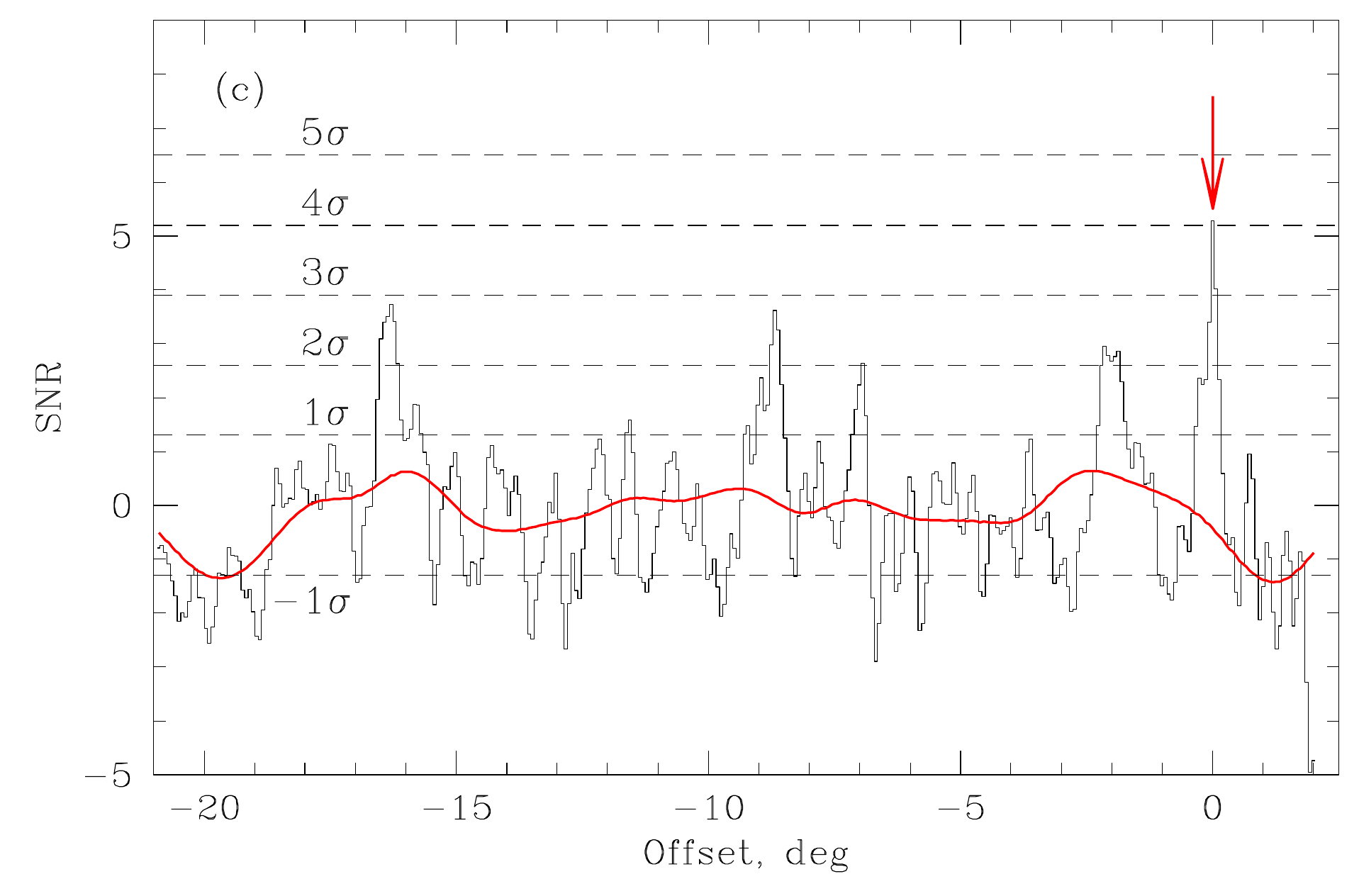}
}
\centerline{
 \includegraphics[width=0.45\textwidth]{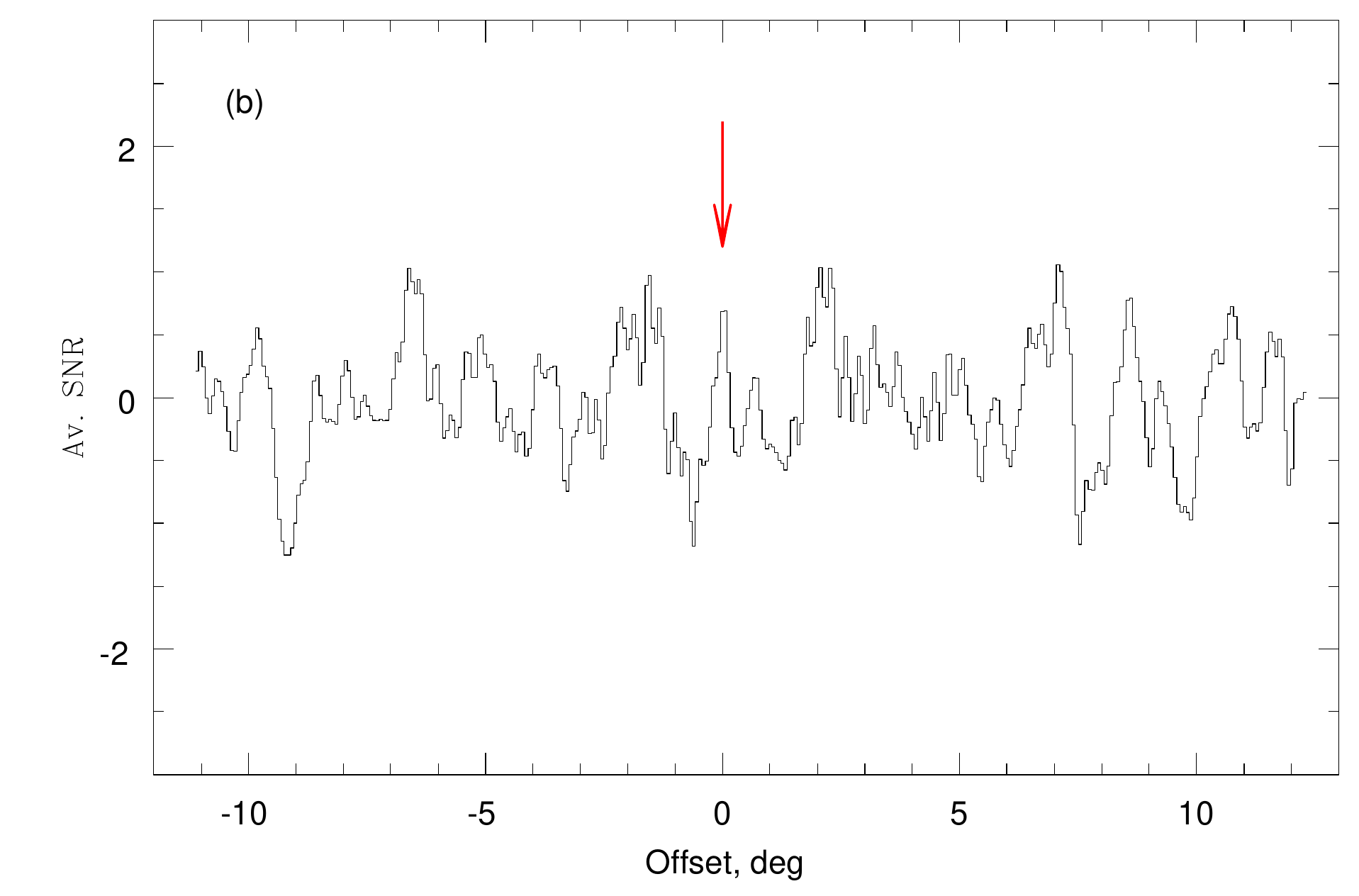}
 \includegraphics[width=0.45\textwidth]{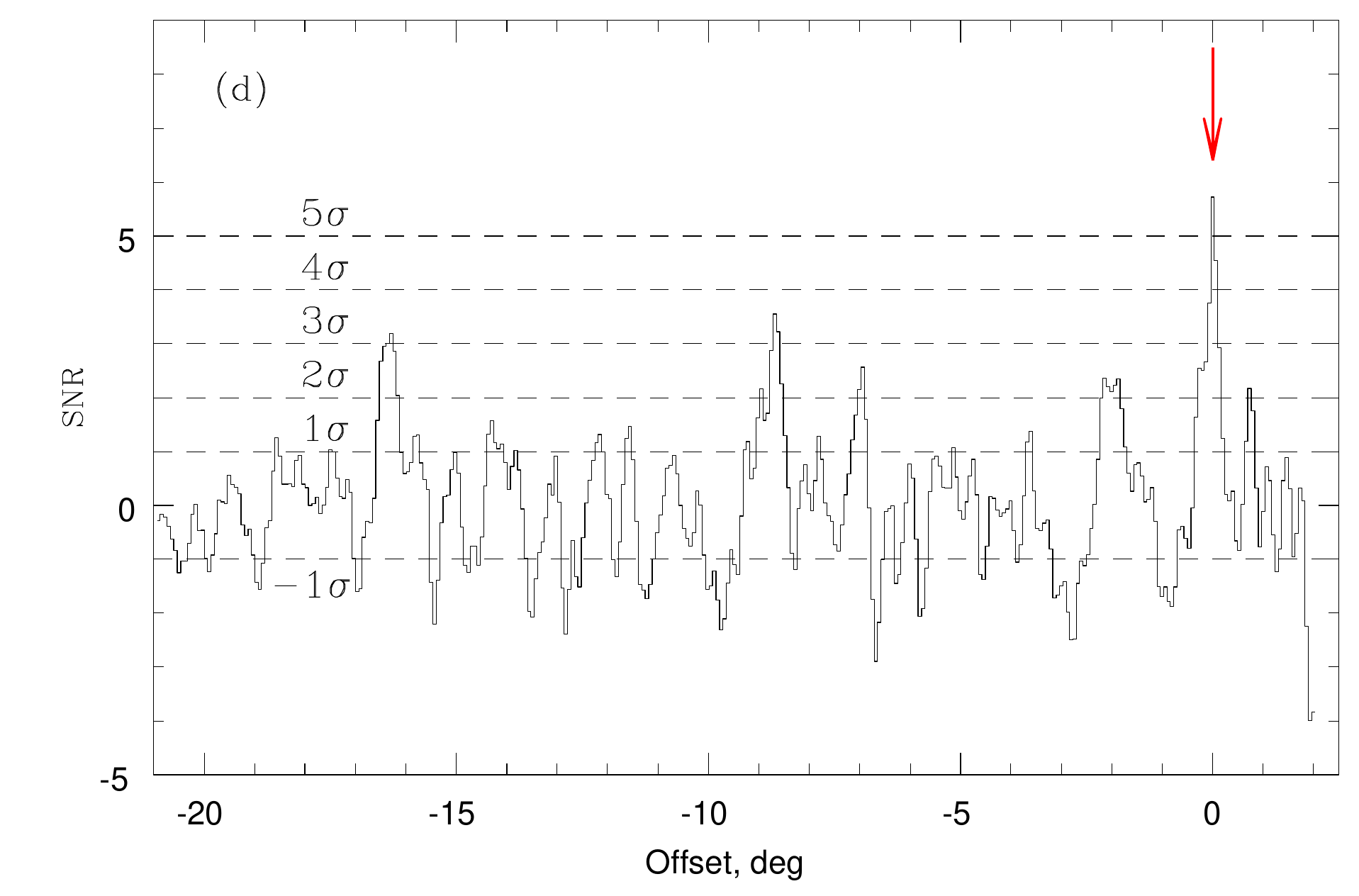}
}
\caption{Illustration of a non-parametric approximation of sky
  background (a and b) and source detection (c and d) with an
  \textit{\`a~trous}\/ DWT algorithm. The black histogram in the spanel
  (a) is a vertically averaged profile of the sky image in
  Fig.~\ref{fig:ngc4151}, extracted from the denoted green
  region. Blue, green and magenta curves are vertical profiles of
  corresponding wavelet components at the spatial scales $J=5,6,7$
  ($\sim1,2,4$ degrees). The sum of these components (red curve)
  represents sky image background approximation which is to be
  subtructed. The corresponding residual profile is shown in the
  bottom panel (b). The abscissa axis is measured in the offset from
  the position of newly detected hard X-ray source IGR~J11203+4531
  (red arrow). The right panels (c and d) are made in the same
  way. The black histogram in the top panel (c) demonstrates an actual
  $S/N$ pixel values extracted along the line perpendicular to the
  green rectangular region in Fig.~\ref{fig:ngc4151}, and crossing the
  source IGR~J11203+4531. The long dashed lines represent the actual
  detection thresholds in standard deviations scaled from the $RMS$
  value measured on all the image pixels}\label{fig:ngc4151:demowv}
\end{figure*}

\subsection{Galactic Center region}
\label{section:atrous:gc}

However, we are mainly interested in improving sensitivity in the
region of the Galactic Plane where most of the exposures were
collected. The sky image of the GP with the maximum available exposure
($\sim20$~Ms in GC) produced by the general reconstruction method is
shown on the upper image in Fig.~\ref{fig:gplane}. The sky background
behind the bright sources is contaminated by strong systematics which
significantly limits sensitivity for source detection. In the same
data set, by taking into account the Galactic X-ray Background and
using DWT sky filtering we obtain a new deep image of the GP which is
demonstrated in the lower panel. As seen from the sky image, most of
the systematic artifacts are removed, leaving a more or less uniform
sky background. Obviously, the quality of the reconstructed sky is
improved.

To demonstrate the efficiency of the improved reconstruction method we
built the SNR distribution of the source-free
$30^{\circ}\times30^{\circ}$ region around the Galactic Center, shown
in Fig.~\ref{fig:sigdistr_gc}. The red and blue histograms represent
general and DWT corrected sky image backgrounds, respectively. The SNR
distribution in the former case has wide non-Gaussian sidelobes. The
blue histogram representing the cleaned sky is narrow, but still far
from normal distribution. This means that systematic artifacts are
reduced, but still present on the sky. We measured the standard
deviation of image pixels masking out bright sources. For the cleaned
and general sky standard deviation relates as $1.33/1.84$, which gives
us the total sensitivity improvement of $\sim28\%$. Taking out the
irreducible Poisson statistics having unit standard deviation, the
suppression of systematic noise in the $\sim20$~Ms deep field of
Galactic Center region is $\sim44\%$.


\begin{figure*}[t]
\includegraphics[width=\textwidth]{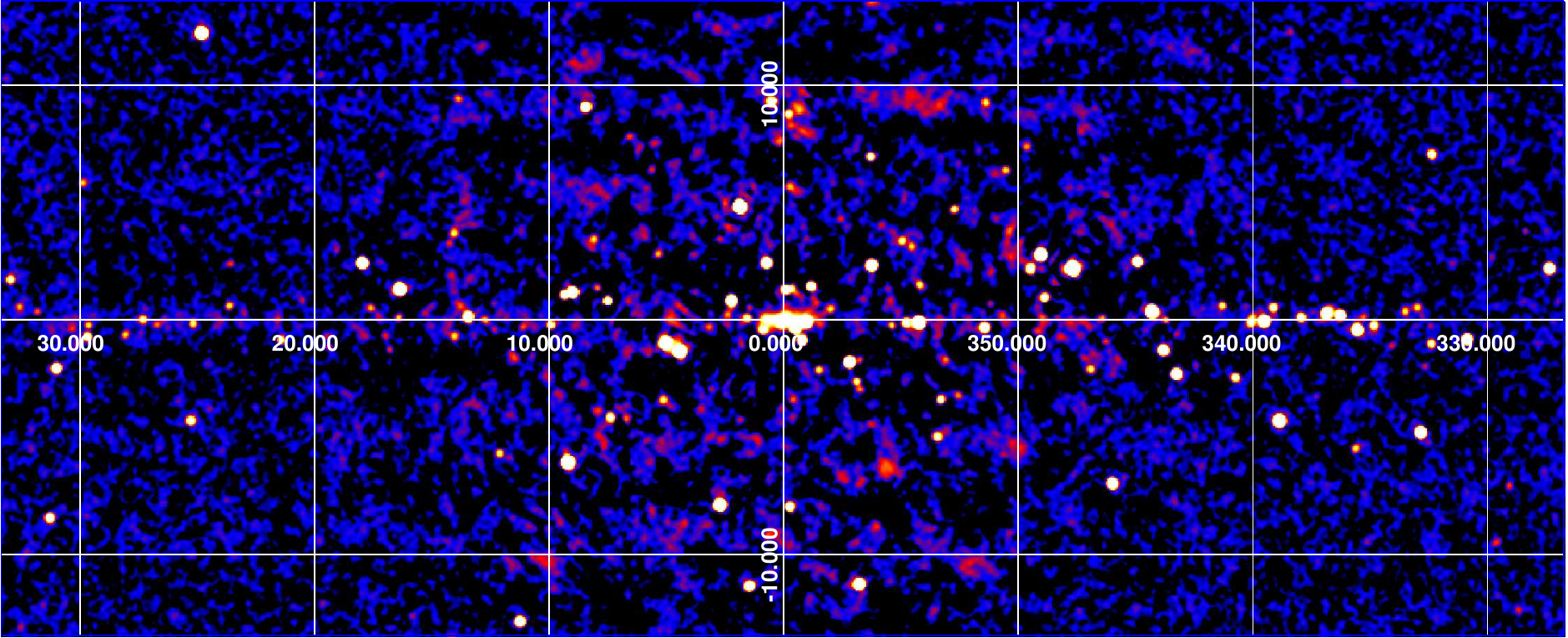}
\includegraphics[width=\textwidth]{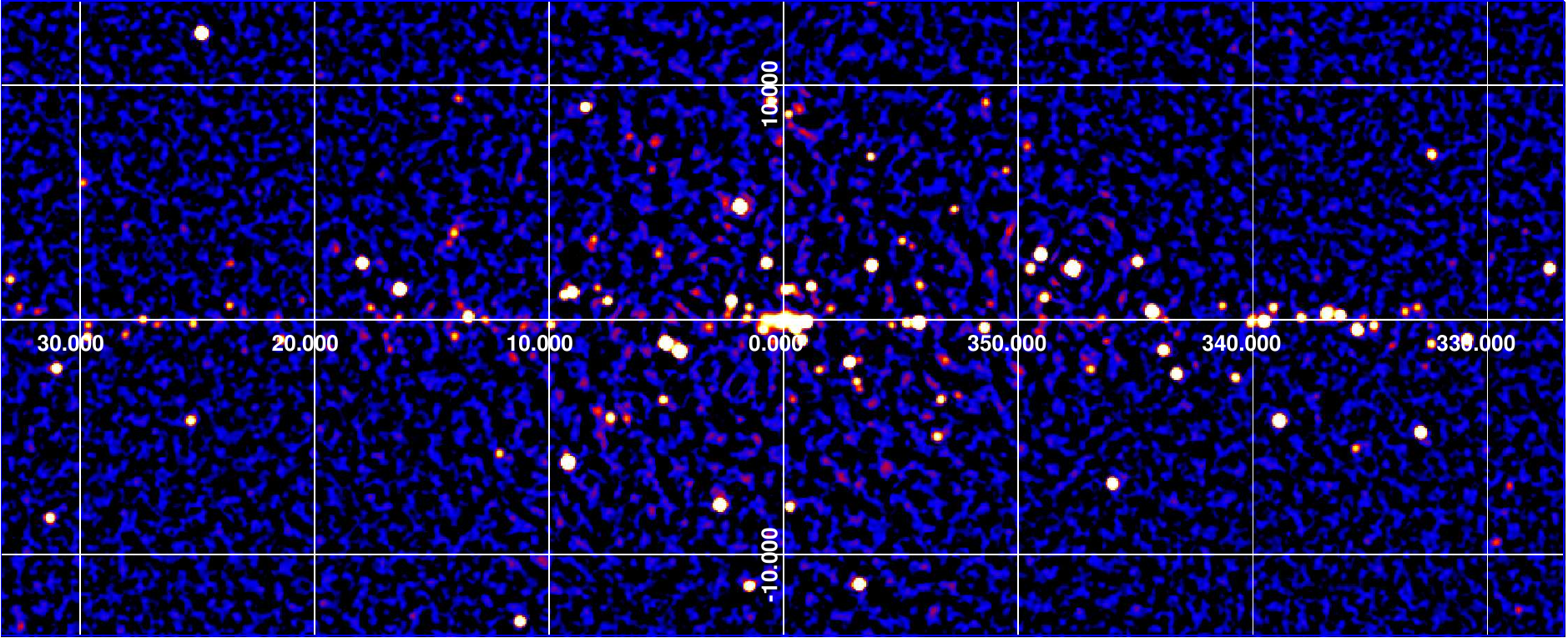}
\caption{Map of the sky region near the Galactic plane obtained with
  IBIS/ISGRI in the $17-60$ keV energy band. The total exposure is
  about 20~Ms in the region of the Galactic Center. \textit{Upper
    panel:} sky mosaic acquired by the general sky reconstruction
  method (see Sect.~\ref{section:method}). \textit{Lower panel:} sky
  mosaic produced by an improved reconstruction algorithm
  (Sect.~\ref{section:method2}). The corresponding signal-to-noise
  distributions of pixels in a $30^{\circ}\times30^{\circ}$ region
  around the GC are shown in Fig.~\ref{fig:sigdistr_gc}.}\label{fig:gplane}
\end{figure*}

\section{Survey}
\label{section:survey}
For our analysis we used all data publically available in July 2009
and observations performed as a part of the GRXE study program (PI
Sunyaev). The latter is mainly based on the Russian quota of INTEGRAL
observing time.

The sky image of any individual observation was produced 
by the modified sky reconstruction method described in this work.
The obtained sky images were added to the all-sky  mosaics covering the whole sky.

\begin{figure}
 \includegraphics[width=0.5\textwidth]{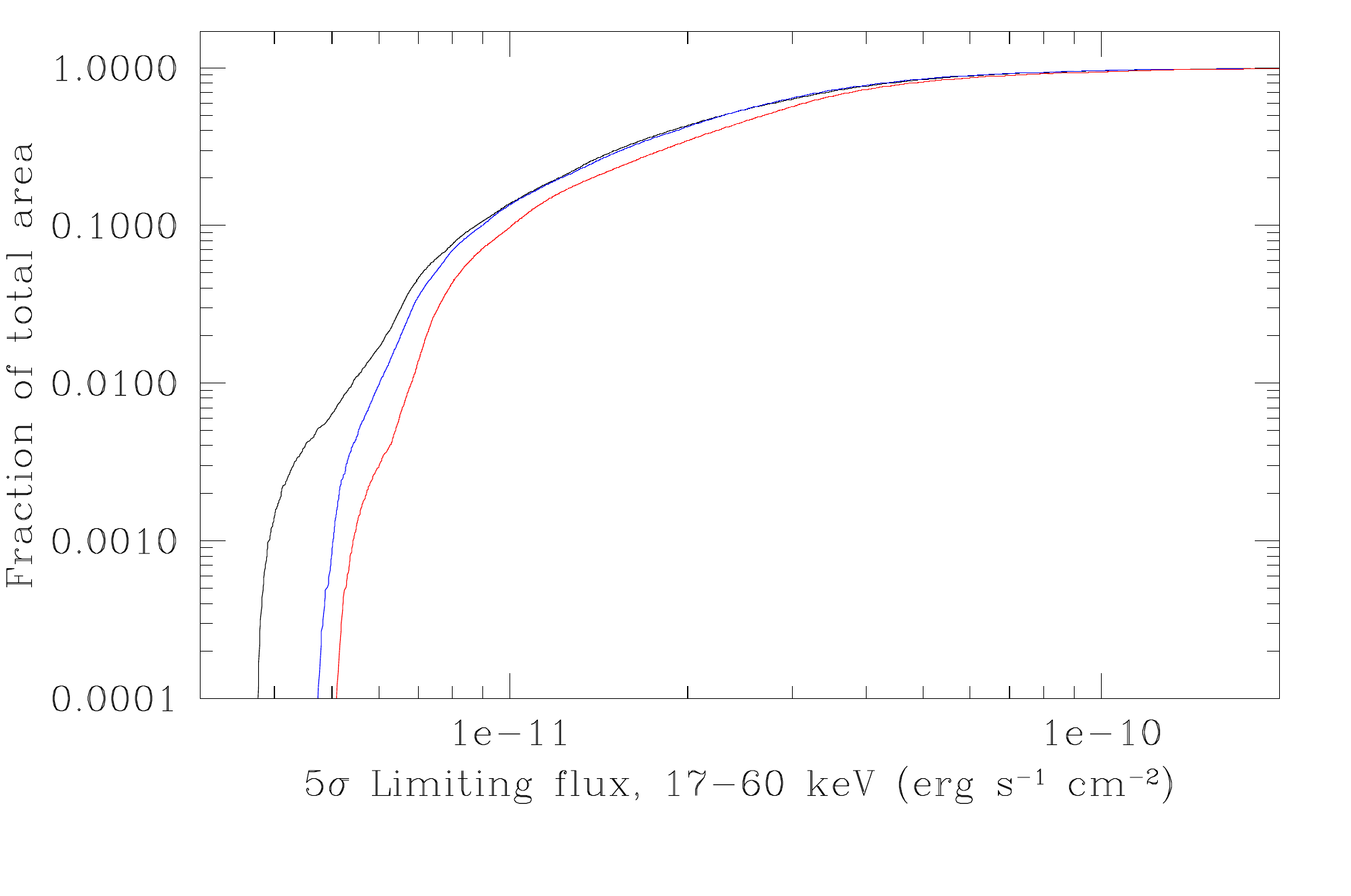}
\caption{Fraction of the sky surveyed as a function of the limiting
  flux for source detection with $5\sigma$ significance. The black
  curve demonstrates sky coverage for nominal sensitivity. The
  effective sensitivity estimated for general and modified sky
  reconstruction methods are shown by the red and blue curve,
  respectively (see Sect.~\ref{section:survey}).}\label{fig:area}
\end{figure}

\begin{figure}
 \includegraphics[width=0.5\textwidth]{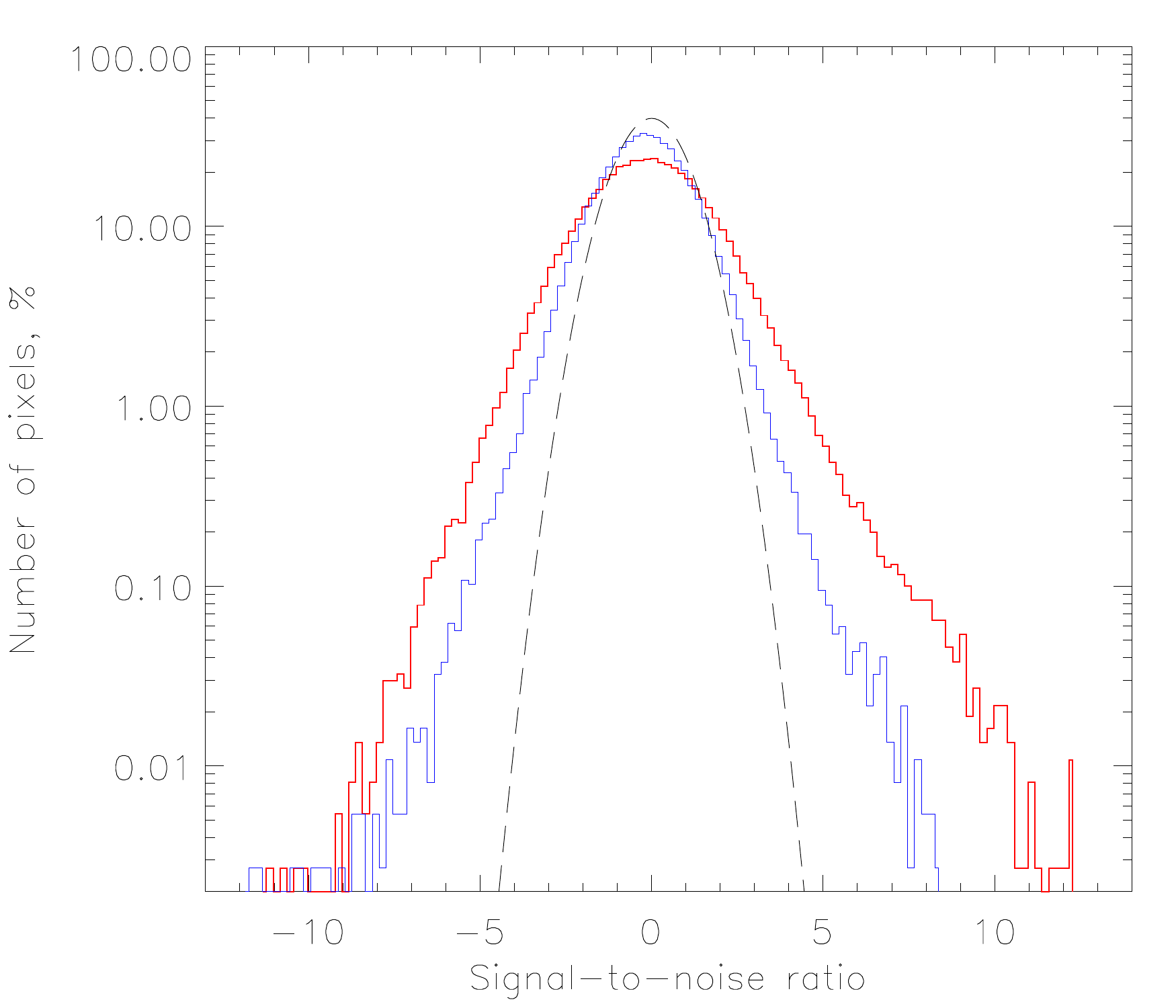}
\caption{Distribution of formal pixel significance in a
  $30^{\circ}\times30^{\circ}$ region around the Galactic Center
  (Fig.~\ref{fig:gplane}). For plot description see
  Fig.~\ref{fig:sigdistr_ngc4151}.}\label{fig:sigdistr_gc}
\end{figure}

\begin{figure}\centerline{
 \includegraphics[width=0.5\textwidth]{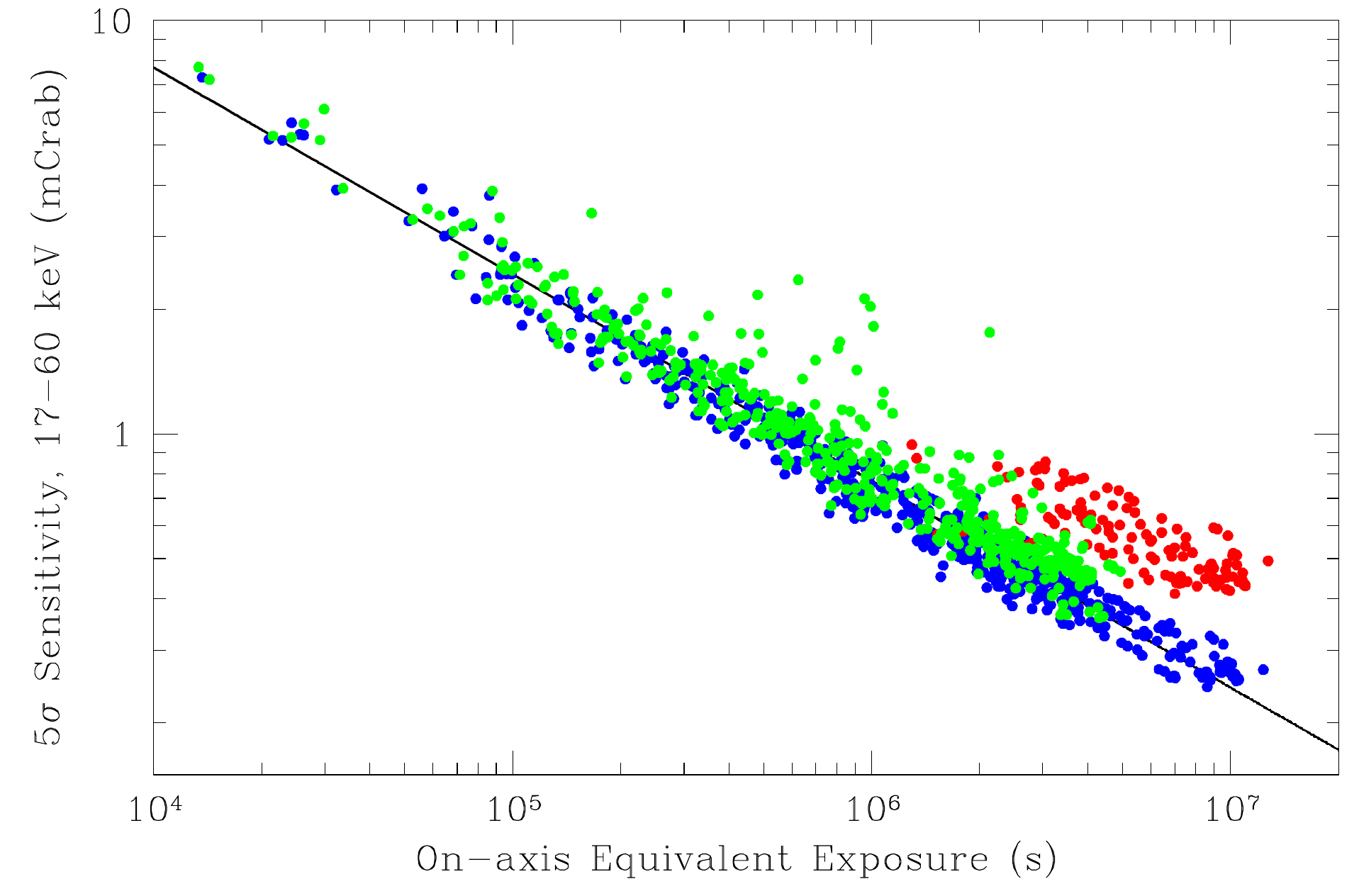}}
\centerline{
 \includegraphics[width=0.5\textwidth]{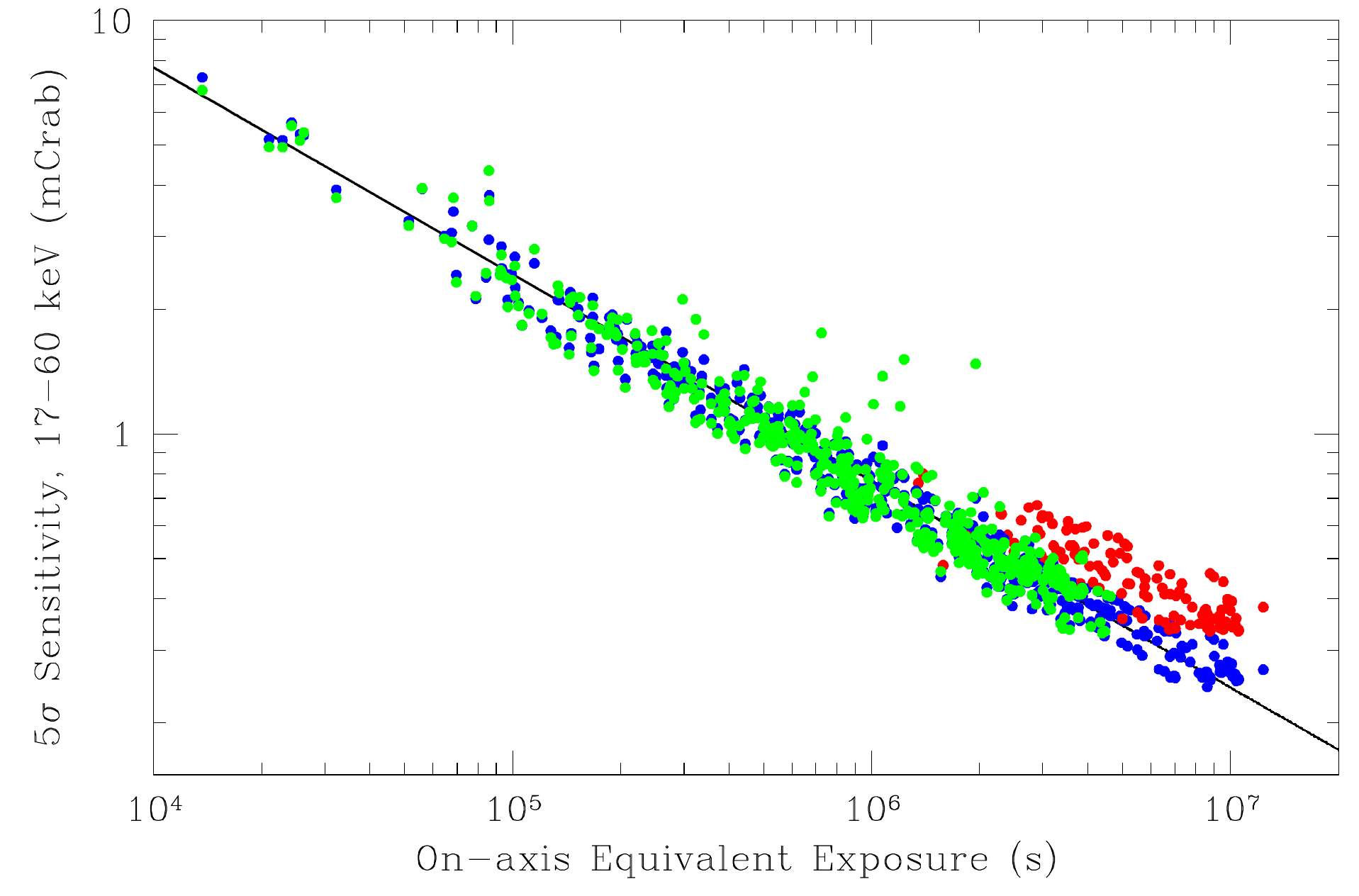}}
\caption{Measured $5\sigma$ limiting flux for catalogued source
  positions on the final total average all-sky map, as a function of
  effective dead-time corrected exposure time. The top and bottom
  panels demonstrate the sensitivity of general and modified sky
  reconstruction methods, respectively. The blue points reflect
  nominal sensitivity changing with exposure by $T^{-1/2}$ as
  expected. The solid line represents the fit to the nominal
  sensitivity versus exposure:
  $f^{5\sigma}_{lim}=0.77\times(T/Ms)^{-1/2}$~mCrab. The green and
  red points are $5\sigma$ actual sensitivity taking into account
  background variations in the field of each
  $20^{\circ}\times20^{\circ}$ sky projection
  (Sect.~\ref{section:survey}). The sensitivity measured in (and out of)
  sky region $|l|<20^{\circ}$ and $|b|<15^{\circ}$ is shown in red
  (green).}\label{fig:sensitivity}
\end{figure}

The survey coverage area is shown in Fig.~\ref{fig:area}. We
calculated a fraction of the sky area covered by the survey at nominal
and effective sensitivity. In the first case, the sensitivity is
essentially the detection threshold, estimated from actual exposure
(black curve). The effective sensitivity was estimated by multiplying
the nominal error map by variances of local background. The general
(Sect.~\ref{section:method}) and modified sky reconstruction method
(Sect.~\ref{section:method2}), are represented by the red and blue
curves, respectively. The limiting flux and sensitivity for $10\%$ and
$90\%$ of the sky coverage are summarized in Table~\ref{tab:flux}.

\textit{General sky reconstruction method:} The survey{'}s limiting flux
for the longest exposure is $\sim35\%$ higher than nominal. For a
large coverage area at high limiting flux, which is typical for
high-latitude observations, the effective sensitivity is $\sim15\%$
lower than expected.

\textit{Modified sky reconstruction method:} The survey effective
coverage is apparently closer to the nominal. In the region of low
limiting flux, the effective sensitivity is $\sim27\%$ worse than
nominal. The sky coverage at high flux is consistent with that expected
from an actual exposure, i.e. Poisson statistics.

In Fig.\ref{fig:sensitivity} we show survey sensitivity as a
function of exposure time. Generally, the survey sensitivity grows
with exposure by $T^{-1/2}$ as expected. The region of high exposure
in the Galactic Center is noticeable by its reduced sensitivity (red dots) in
contrast with extragalactic observations (green dots) where the limiting
flux is expected to follow pure statistics (blue dots). Again, the
survey sensitivity of the modified method is closer to a nominal sensitivity.

\begin{table}
\begin{center}
\caption{Survey limiting flux in mCrab\textit{s} for source
  detection at the $5\sigma$ significance level (17-60~keV, $1\textrm{
    mCrab}=1.43\times10^{-11}$\ergscm).}
\label{tab:flux}
\begin{tabular}{lccc}
\hline 
\hline 
\noalign{\smallskip} Category & Nominal & Gen. method & Mod. method \\ 
\noalign{\smallskip}\hline\noalign{\smallskip} \hline
     Lim. flux & 0.26 &  0.35 & 0.33 \\
     10\% sky & 0.60 &  0.70 & 0.63 \\
     90\% sky & 4.32 &  5.00 & 4.32 \\
\hline
\noalign{\smallskip}\hline\noalign{\smallskip}
\end{tabular}
\end{center}
\end{table}

\section{Conclusion}
\label{section:end}

We presented the improved sky reconstruction method for the IBIS
telescope which suppresses the systematic effects. Firstly, the method
takes into account extended Galactic X-ray Ridge emission which
strongly affects the background illumination of the ISGRI
detector. Secondly, we applied a non-parametric (model-free)
background approximation based on an \textit{\`a~trous}\/ wavelet
decomposition. The wavelet cleaning was naturally integrated into the
sky reconstruction process with the main advantage that we knew
exactly what we were filtering out without distorting the original sky
flux. The overall systematic noise in the $\sim20$~Ms deep field of
the Galactic Center was reduced by $\sim44\%$ improving the the total
sensitivity of observations by $\sim28\%$. The reconstructed sky
images of high galactic latitude fields were practically free from
systematic residuals and the sensitivity was consistent with that
expected from Poisson statistics.

Most of the INTEGRAL observing time was spent in the Galactic Plane
and Center, giving us the possibility to conduct the most sensitive
survey ever made of the Milky Way above 20 keV. The minimal detectable
flux with a $5\sigma$ detection level reached the level of
$3.7\times10^{-12}$~\ergscm, which is $\sim0.26$~mCrab in the
17-60~keV energy band. The survey covered $90\%$ of the sky down to
the flux limit of $6.2\times10^{-11}$~\ergscm ~($\sim4.32$~mCrab) and
$10\%$ of the sky area down to the flux limit of
$8.6\times10^{-12}$~\ergscm ~($\sim0.60$~mCrab). A catalogue of sources
detected in the survey is presented in paper by Krivonos et al.,
(2010b, in prep.).

\begin{acknowledgements}
We are grateful to the anonymous referee for the critical remarks
which helped us improve the paper. The data used were obtained from
the European and Russian INTEGRAL Science Data
Centers\footnote{http://isdc.unige.ch}'\footnote{http://hea.iki.rssi.ru/rsdc}.
The work was supported by the President of the Russian Federation
(through the program supporting leading scientific schools, project
NSH-5069.2010.2), by the Presidium of the Russian Academy of
Sciences/RAS (the program ``Origin, Structure, and Evolution of
Objects of the Universe''), by the Division of Physical Sciences of
the RAS (the program ``Extended objects in the Universe'', OFN-16),
and by the Russian Basic Research Foundation (project 09-02-00867). SS
acknowledges the support of the Dynasty Foundation. RK would like to
express his sincere gratitude to Kate O{'}Shea for her support in the
preparation of the paper.

\end{acknowledgements}


\begin{thebibliography}{}

\bibitem[Barthelmy et al.(2005)]{bat} Barthelmy, S.~D., et al.\ 2005, Space Science Reviews, 120, 143 
\bibitem[Bassani et al.(2006)]{bassani06} Bassani, L., et al.\ 2006, \apjl, 636, L65
\bibitem[Bazzano et al.(2006)]{bazzano06} Bazzano, A., et al.\ 2006, \apjl, 649, L9 
\bibitem[Beckmann et al.(2009)]{beckmann09} Beckmann, V., et al.\ 2009, \aap, 505, 417 
\bibitem[Bird et al.(2004)]{birdI} Bird, A.~J., et al.\ 2004, \apjl, 607, L33
\bibitem[Bird et al.(2006)]{birdII} Bird, A.~J., et al.\ 2006, \apj, 636, 765
\bibitem[Bird et al.(2007)]{birdIII} Bird, A.~J., et al.\ 2007, \apjs, 170, 175
\bibitem[Bird et al.(2010)]{birdIV} Bird, A.~J., et al.\ 2010, \apjs, 186, 1 
\bibitem[Biviano et al.(1996)]{biviano96} Biviano, A., Durret, F., Gerbal, D., Le Fevre, O., Lobo, C., Mazure, A., \& Slezak, E.\ 1996, \aap, 311, 95
\bibitem[Cusumano et al.(2010)]{palermo36} Cusumano, G., et al.\ 2010, \aap, 510, A48 
\bibitem[Del Santo et al.(2003)]{delsanto03} Del Santo, M., et al.\ 2003, \aap, 411, L369 
\bibitem[Dwek et al.(1995)]{dwek95} Dwek E., et al., 1995, ApJ, 445, 716
\bibitem[Eckert et al.(2008)]{eckert2008} Eckert, D., Produit, N., Paltani, S., Neronov, A., \& Courvoisier, T.~J.-L.\ 2008, \aap, 479, 27 
\bibitem[Fenimore \& Cannon (1978)]{fenimore78} Fenimore, E.E., Cannon, T.M.: 1978, APO, 17, 337
\bibitem[Fenimore \& Cannon (1981)]{fenimore81} Fenimore, E.~E., Cannon T.~M., \ 1981 Applied Optics, 20, 1858.
\bibitem[Gehrels et al.(2004)]{swift} Gehrels, N., et al.\ 2004, \apj, 611, 1005 
\bibitem[Goldwurm et al.(2003)]{goldwurm03} Goldwurm A., et al., 2003, A\&A, 411, L223 
\bibitem[Grebenev et al.(1995a)]{grebenev95} Grebenev, S.~A., Forman, W., Jones, C., \& Murray, S.\ 1995, \apj, 445, 607 
\bibitem[Grebenev et al.(1995b)]{grebenev95wv} Grebenev, S.~A., Pavlinsky, M.~N., \& Sunyaev, R.~A.\ 1995, Proc.~of the Workshop ''Imaging in High Energy Astronomy'' (held in Anacapri, Sept.~26-30, 1994, eds.~Bassani, L.; di Cocco, G.), Kluwer Academic Publishers, Dordrecht, pp.~155-158, 155 
\bibitem[Krivonos et al.(2005)]{krietal05} Krivonos, R., Vikhlinin, A., Churazov, E., Lutovinov, A., Molkov, S., \& Sunyaev, R.\ 2005, \apj, 625, 89
\bibitem[Krivonos et al.(2007a)]{krietal07a} Krivonos, R., Revnivtsev, M., Churazov, E., Sazonov, S., Grebenev, S., \& Sunyaev, R.\ 2007, \aap, 463, 957
\bibitem[Krivonos et al.(2007b)]{krietal07b} Krivonos, R., Revnivtsev, M., Lutovinov, A., Sazonov, S., Churazov, E., \& Sunyaev, R.\ 2007, \aap, 475, 775
\bibitem[Lebrun(2005)]{lebrun05} Lebrun, F.\ 2005, IEEE Transactions on Nuclear Science, 52, 3119 
\bibitem[Molkov et al.(2004)]{molkov2004} Molkov S.~V., Cherepashchuk A.~M., Lutovinov A.~A., Revnivtsev M.~G., Postnov K.~A., Sunyaev R.~A., 2004, AstL, 30, 534
\bibitem[Reglero et al.(2001)]{reglero01} Reglero V., et al., 2001, ESASP, 459, 619 
\bibitem[Revnivtsev et al.(2003)]{revetal03a} Revnivtsev, M.~G., Sazonov, S.~Y., Gilfanov M.~R., Sunyaev, R.~A.\ 2003a, Astronomy Letters, 29, 587
\bibitem[Revnivtsev et al.(2004)]{revetal04} Revnivtsev, M., Sunyaev, R., Varshalovich, D., et al.\ 2004, Astronomy Letters, 30, 382
\bibitem[Revnivtsev et al.(2006a)]{revetal06a} Revnivtsev, M.~G., Sazonov, S.~Y., Molkov, S.~V., Lutovinov, A.~A., Churazov, E.~M., \& Sunyaev, R.~A.\ 2006, Astronomy Letters, 32, 145
\bibitem[Revnivtsev et al.(2006b)]{revetal06b} Revnivtsev, M., Sazonov, S., Gilfanov, M., Churazov, E., \& Sunyaev, R.\ 2006, \aap, 452, 169
\bibitem[Rodriguez et al.(2003)]{rodriguez03} Rodriguez, J., et al.\ 2003, \aap, 411, L373 
\bibitem[Rosati et al.(1995)]{rosati95} Rosati, P., della Ceca, R., Burg, R., Norman, C., \& Giacconi, R.\ 1995, \apjl, 445, L11 
\bibitem[Sazonov et al.(2007)]{sazonov07} Sazonov, S., Revnivtsev, M., Krivonos, R., Churazov, E., \& Sunyaev, R.\ 2007, \aap, 462, 57
\bibitem[Skinner et al.(1987)]{skinner87} Skinner, G.~K., Ponman, T.~J., Hammersley, A.~P., \& Eyles, C.~J.\ 1987, \apss, 136, 337
\bibitem[Slezak et al.(1994)]{slezak94} Slezak, E., Durret, F., \& Gerbal, D.\ 1994, \aj, 108, 1996 
\bibitem[Starck \& Murtagh(1994)]{starck94} Starck, J.-L., \& Murtagh, F.\ 1994, \aap, 288, 342
\bibitem[Tueller et al.(2009)]{bat22} Tueller, J., et al.\ 2009, arXiv:0903.3037 
\bibitem[Ubertini et al. (2003)]{ibis} Ubertini P., et al., 2003, A\&A, 411, L131
\bibitem[Vikhlinin et al.(1997)]{vikhlinin96} Vikhlinin, A., Forman, W., \& Jones, C.\ 1997, \apjl, 474, L7 
\bibitem[Winkler et al. (2003)]{integral} Winkler C., et al., 2003, A\&A, 411, L1
\bibitem[Winkler et al.(1999)]{core} Winkler, C., Gehrels, N., Lund, N., Sch{\"o}nfelder, V., \& Ubertini, P.\ 1999, Astrophysical Letters Communications, 39, 361 
\bibitem[Worrall et al. (1982)]{worrall82} Worrall D.~M., Marshall F.~E., Boldt E.~A., Swank J.~H., 1982, ApJ, 255, 111


\end{thebibliography}
\end{document}